\documentclass[aps,prx,longbibliography,reprint,twocolumn,superscriptaddress,floatfix,nofootinbib]{revtex4-2}
\usepackage{epsfig,amsmath,amssymb,color,comment,physics, dsfont, bbold}
\usepackage[makeroom]{cancel}
\usepackage[caption=false]{subfig}
\usepackage{mathrsfs}
\usepackage[normalem]{ulem}
\usepackage[english]{babel}
\usepackage{dsfont}
\usepackage{float}
\usepackage{color}
\usepackage{graphicx} 
\usepackage[dvipsnames]{xcolor}
\usepackage{tikz}
\usepackage[none]{hyphenat}
\usepackage{bm}
\usepackage{placeins}
\newcommand{\defaulttensorsize}{10pt}
\newcommand{\tensorsize}{\defaulttensorsize}
\usetikzlibrary{shapes.geometric}
\usetikzlibrary{shapes.misc}
\tikzstyle{tensor}=[draw, inner sep=0, outer sep=0, minimum size=\tensorsize]
\tikzstyle{notensor}=[inner sep=0, outer xsep=2pt, outer ysep=0, minimum size=\tensorsize]
\tikzstyle{atensor}=[tensor, circle]
\tikzstyle{ctensor}=[tensor, circle]
\tikzstyle{dtensor}=[tensor, diamond]
\tikzstyle{wtensor}=[tensor]
\tikzstyle{ltensor}=[tensor, rounded rectangle, rounded rectangle left arc=none]
\tikzstyle{rtensor}=[tensor, rounded rectangle, rounded rectangle right arc=none]
\tikzstyle{etensor}=[tensor, minimum height=(1cm/\defaulttensorsize*0.5*2+1)*\tensorsize]
\tikzstyle{widetensor}[2]=[tensor, minimum width=(1cm/\defaulttensorsize*0.75*(#1-1)+1)*\tensorsize]

\tikzstyle{tensornetwork}=[baseline=-0.25em, xscale=0.75, yscale=0.5,
                           every node/.style={inner sep=0, outer xsep=2pt, outer ysep=5pt, text depth=0pt},
                           execute at end picture={\path (current bounding box.south west) +(-2pt, 0) (current bounding box.north east) +(2pt, 0);}]

\usetikzlibrary{quantikz}

\usepackage[bookmarks=true,colorlinks,linkcolor=blue,urlcolor=NavyBlue,citecolor=RoyalBlue]{hyperref}
\newcommand{\beginsupplement}{%
    \setcounter{table}{0}
    \renewcommand{\thetable}{S\arabic{table}}%
    \setcounter{figure}{0}
    \renewcommand{\thefigure}{S\arabic{figure}}%
    \setcounter{equation}{0}
    \renewcommand{\theequation}{S\arabic{equation}}%
    \setcounter{section}{0}
    \renewcommand{\thesection}{S\arabic{section}}%
    \setcounter{page}{1}
   }

\begin{document}
\sloppy{}
\title{
Quantum many-body mixed phase space revealed by hybrid feedback control
}

\newcommand{\zju}{School of Physics, ZJU-Hangzhou Global Scientific and Technological Innovation Center, \\and Zhejiang Key Laboratory of Micro-nano Quantum Chips and Quantum Control,Zhejiang University,Hangzhou 310000,China}
\newcommand{\leeds}{School of Physics and Astronomy, University of Leeds, Leeds LS2 9JT,  United Kingdom}

\author{Hang Dong} 
\thanks{These authors contributed equally.}
\affiliation{\zju}

\author{Jie Ren}
\thanks{These authors contributed equally.}
\affiliation{\leeds}

\author{Andrew Hallam} 
\thanks{These authors contributed equally.}
\affiliation{\leeds}

\author{Han Wang} 
\affiliation{\zju}
\author{Zhengyi Cui}
\affiliation{\zju}
\author{Yiren Zou}
\affiliation{\zju}
\author{Junlin Wang}
\affiliation{\zju}
\author{Hekang Li}
\affiliation{\zju}
\author{Qiujiang Guo}
\affiliation{\zju}
\author{Zhen Wang}
\email{2010wangzhen@zju.edu.cn}
\affiliation{\zju}
\author{Lei Ying}
\email{leiying@zju.edu.cn}
\affiliation{\zju}
\author{Zlatko Papi\'c}\email{z.papic@leeds.ac.uk}
\affiliation{\leeds}

\begin{abstract}
Understanding how complex systems transition between order and chaos is a central challenge of nonequilibrium physics. 
While weak perturbations of classical integrable systems give rise to a mixed phase space of coexisting regular and chaotic trajectories, analogous behavior in interacting quantum many-body systems has remained elusive.
Here we develop and experimentally implement a hybrid quantum–classical feedback protocol that autonomously discovers and stabilizes long-lived regular trajectories in a superconducting quantum processor. Each iteration combines short-time quantum evolution with classical optimization that projects the dynamics back onto a low-entanglement variational manifold, effectively distilling coherence from chaotic evolution. The stabilized trajectories reveal a quantum many-body mixed phase space emerging from nonlinear variational dynamics, without a direct analogue in classical or few-body quantum systems. Our results establish a versatile framework for algorithmic discovery and control of coherent dynamics previously inaccessible to experiment.

\end{abstract}

\maketitle

The coexistence of regular and chaotic motion is a universal feature of dynamical systems, ranging from  planetary orbits and ocean waves to turbulence and transport in complex materials~\cite{StrogatzBook,lichtenberg2013regular}. In classical mechanics, this interplay is governed  by the Kolmogorov–Arnold–Moser (KAM) theorem~\cite{arnol2013mathematical}, which shows that weak perturbations of integrable systems preserve a finite measure of invariant tori, producing a \emph{mixed phase space} where regular and chaotic trajectories coexist. In the quantum realm, remnants of integrability can survive in few-body systems described by a single collective degree of freedom~\cite{Neill2016} or under weak perturbations of models solvable by the Bethe ansatz~\cite{GlimmersKAM}. However, strongly-interacting quantum systems can exhibit qualitatively new forms of dynamics that resemble mixed phase space but have no direct few-body analog. In such systems, the coexistence of coherent and chaotic motion may arise from the collective evolution of entangled degrees of freedom rather than from a conventional classical limit $\hbar \to 0$~\cite{Bohigas1993}.

The process of thermalization in generic quantum systems is governed by the eigenstate thermalization hypothesis (ETH)~\cite{DeutschETH,SrednickiETH}, which states that highly excited eigenstates reproduce the thermal expectation values of local observables. Systems that weakly break the ETH, e.g., those with quantum many-body scars and Hilbert-space fragmentation~\cite{Serbyn2021,MoudgalyaReview,ChandranReview}, are a natural starting point to search for quantum many-body analogs of mixed phase space as these systems are generally chaotic yet host special nonthermalizing states~\cite{Bernien2017,Turner2017}.
However, such initial states are exceedingly rare in the many-body Hilbert space and thus difficult to identify. A framework capable of revealing and stabilizing nonthermal trajectories is therefore needed to access new universality classes of nonergodic quantum dynamics beyond currently known ETH violations.

Here we demonstrate a hybrid quantum–classical feedback protocol that autonomously discovers and stabilizes coherent trajectories in interacting many-body systems. Each iteration of the protocol alternates between short-time quantum evolution and classical optimization that steers the system back toward a low-entanglement variational manifold~\cite{scarfinder}. Repeated application of this evolution–projection cycle filters out the entangling behavior of chaotic motion, converging to a stable periodic orbit if it exists. Conceptually, the method blends the time-dependent variational principle (TDVP)~\cite{kramer1981geometry,Haegeman,wenwei18TDVPscar,Hallam2019} with active feedback control~\cite{Boscain2021,Duncan2025} by transforming the geometric projection of quantum dynamics into a practical measurement-based algorithm, suitable for current quantum hardware.

We implement the hybrid feedback scheme on a programmable superconducting-qubit processor that realizes an interacting Su–Schrieffer–Heeger (SSH) ladder~\cite{SuSchriefferHeeger,Zhang2022}. By mapping the ladder's dynamics onto a shallow variational circuit and monitoring local fidelity, we uncover islands of regular motion within a chaotic sea utilizing a many-body analog of the Poincar\'e section. The resulting trajectories reveal a \emph{quantum many-body mixed phase space}, which was previously proposed to occur in Rydberg atom arrays~\cite{Michailidis2020} but has so far eluded experimental observation. This emergent structure originates from nonlinear TDVP dynamics and many-body correlations rather than the conventional classical limit, serving as an emergent semiclassical phase-space picture that bridges classical chaos and quantum ergodicity breaking.

\begin{figure*}[t]
    \centering   
    \includegraphics[width=0.95\linewidth]{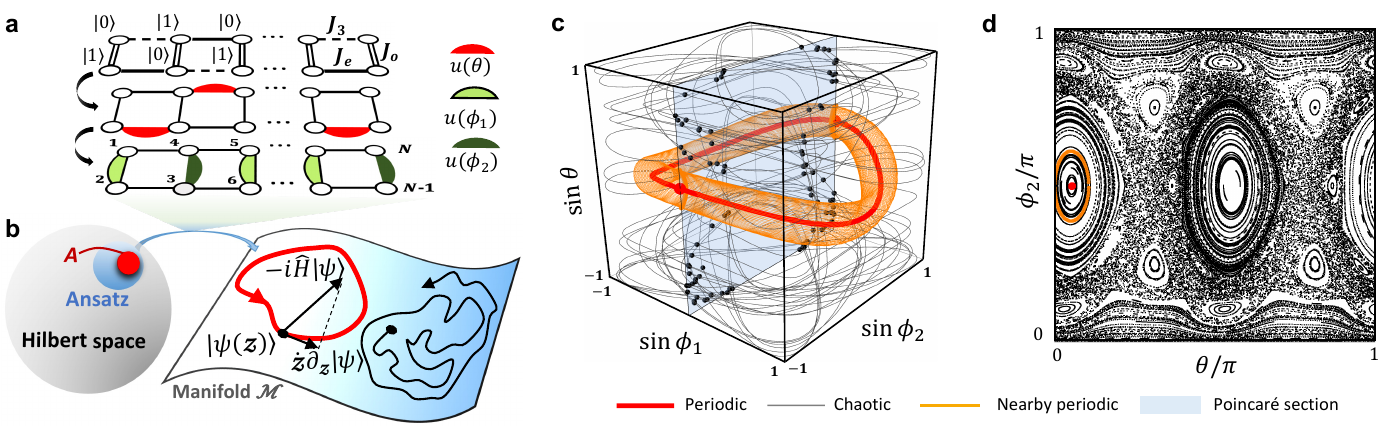}
    \caption{ {\bf Superconducting circuit with a mixed phase space.}
    \textbf{a} The shallow variational circuit used to approximate the real-time dynamics of an interacting SSH  ladder, Eq.~(\ref{eq:Hamiltonian}), on our superconducting qubit processor. The inter-leg coupling $J_o$ is uniform, while the intra-leg couplings alternate between $J_e$ and $J_3$, defining an SSH-type dimerization pattern. The  calibrated device couplings  are $J_e/2\pi=3.0~\mathrm{MHz}$, $J_3/2\pi=0.25~\mathrm{MHz}$, and $J_o/2\pi=5.0~\mathrm{MHz}$. 
    Starting from the N\'eel-ordered product state, two layers of entangling XY gates are applied: first $\hat u(\theta)$ on even bonds, followed by alternating $\hat u(\phi_1)$ and $\hat u(\phi_2)$ on odd bonds (see text).
    \textbf{b} The circuit ansatz from panel {\bf a} [Eq.~\eqref{eq:shallow_circuit}] defines a three-dimensional variational manifold $\mathcal{M}$ (blue region) embedded in the full Hilbert space, parametrized by $\bm{z}=(\theta,\phi_1,\phi_2)$, which hosts a low-dimensional projection of quantum dynamics. The exact Schr\"odinger evolution, $-i\hat{H}|\psi\rangle$, is projected onto the tangent space of $\mathcal{M}$ using TDVP. The red region~$A$ illustrates the subset of ansatz states associated with the stable periodic-orbit island of the TDVP flow. Depending on initial conditions, the projected trajectory may form a regular orbit (red curve) or wander chaotically through the manifold (black curve). 
    \textbf{c} Representative classical trajectories in the reduced parameter space $(\theta,\phi_1,\phi_2)$, obtained by integrating the TDVP equations of motion (see Methods for the explicit expressions). Shown are a stable periodic orbit (red), a nearby regular trajectory (orange), and a chaotic trajectory (gray). The plane $\phi_1=0\bmod 2\pi$ (light blue) marks the slice defining the Poincar\'e section. 
    \textbf{d} Poincar\'e section obtained from the intersections of TDVP trajectories with the plane $\phi_1=0\bmod 2\pi$ in panel {\bf c}. The section is generated from 350 trajectories evolved up to a late time $t = 2000$ (in units $\hbar |J_o|=1$), each initialized from random points on the manifold. Regular trajectories form discrete point sets lying along smooth invariant curves (the cross-sections of tori); two sets of representative points are highlighted in red and orange.  Black points include both chaotic trajectories, which populate extended regions, and unhighlighted regular points on other tori. The section reveals the coexistence of regular islands and a chaotic sea within the TDVP dynamics.
    }
    \label{fig:overview}
\end{figure*}

\section*{Superconducting circuit with mixed phase space}

The versatility of superconducting qubit platforms, including high-fidelity initialization, tunable two-qubit couplings, and local measurements, makes it possible to access both ergodic and nonergodic dynamical regimes of quantum many-body systems, while resolving physical observables at the single-qubit level. In this work, we exploit these capabilities to design an interacting spin-$1/2$ model inspired by the SSH chain~\cite{SuSchriefferHeeger} that hosts a many-body analogue of mixed phase space. Our model is realized on a square ladder with $N=24$ qubits, schematically illustrated in Fig.~\ref{fig:overview}\textbf{a}. The circuit encodes the following Hamiltonian,
\begin{eqnarray}\label{eq:Hamiltonian}
    \nonumber \hat{H} = &-& \sum_{n=1}^{N/2-1}\Big(
    J^t_n \,\operatorname{XY}_{(n,1),(n+1,1)} 
    + J^b_n \,\operatorname{XY}_{(n,2),(n+1,2)}\Big) \\
    &-& \sum_{n=1}^{N/2} J_n^{\mathrm{int}} \,\operatorname{XY}_{(n,1),(n,2)}\,,    
\end{eqnarray}
where $(n,r)$ labels the $n$th qubit in the row $r\in \{1,2\}$ and $\operatorname{XY}_{i,j} \equiv \hat{\sigma}^+_i \hat{\sigma}^-_j + \hat{\sigma}^+_j \hat{\sigma}^-_i$ denotes an exchange interaction in terms of the standard Pauli raising and lowering operators, $\hat\sigma^\pm$. The inter-leg couplings are chosen uniformly, $J^{\mathrm{int}}_n = J_o$, while intra-leg couplings alternate as $J^t_{2n} = J^b_{2n+1} = J_e$ and $J^t_{2n+1} = J^b_{2n} = J_3$. For convenience, we map the ladder circuit onto a one-dimensional (1D) chain that can be conveniently labeled by a single-site index, as indicated in Fig.~\ref{fig:overview}{\bf a} -- a notation we will adopt throughout this paper. 

Different choices of the couplings allow us to smoothly interpolate between integrable and chaotic regimes. For instance, when $J_e = J_3 = 0$, the circuit breaks up into decoupled dimers, which support exactly periodic rung oscillations, such as from a product state $\ket{\varphi}=\ket{01100110\dots}$. When $J_e$ and $J_3$ are nonzero, the system is chaotic, yet the same initial state continues to approximately oscillate, as previously observed in~\cite{Zhang2022}. In the Supplementary Material (SM)~\cite{SOM}, we demonstrate that the dynamics of the model (\ref{eq:Hamiltonian}) are much richer: even in the thermodynamic limit $N{\to}\infty$, the model displays a breakdown of ETH: it hosts continuous families of initial conditions at the same energy density yet with starkly different slopes of entanglement growth. Below we develop a framework that allows us to interpret this breakdown as a many-body manifestation of mixed phase space.

To capture the coexistence of coherent and thermalizing dynamics, we introduce a shallow variational circuit ansatz that serves as an effective low-dimensional representation of the full quantum dynamics. The ansatz consists of two-qubit entangling gates $\hat u_{i,j}(\theta)=\exp(i\theta\,\operatorname{XY}_{i,j})$ applied sequentially along the lattice, starting from a N\'eel-ordered product state. We first apply $\hat u(\theta)$ on all even bonds to form the first layer operator $\hat U_1(\theta)=\hat u_{2,3}(\theta)\hat u_{4,5}(\theta)\cdots$, followed by alternating $\hat u(\phi_1)$ and $\hat u(\phi_2)$ on odd bonds, which define the second layer operator $\hat U_2(\phi_1,\phi_2)=\hat u_{1,2}(\phi_1)\hat u_{3,4}(\phi_2)\dots$, as shown in Fig.~\ref{fig:overview}{\bf a}. The resulting shallow circuit ansatz is
\begin{equation}
    |\psi(\bm z)\rangle \equiv \hat U_{\mathcal M}(\bm z)|0101\dots\rangle, \quad\; 
    \hat U_{\mathcal M}(\bm z) = \hat U_2(\phi_1,\phi_2)\hat U_1(\theta), \label{eq:shallow_circuit}
\end{equation}
where we introduce the collective label for the circuit parameters $\bm z \equiv (\theta,\phi_1,\phi_2)$. This ansatz is motivated by first-order Trotterization of the SSH ladder Hamiltonian without the $J_3$ term, making it quantitatively accurate in the regime $J_e,J_3 \ll J_o$ and for short evolution times. It thus faithfully represents the low-entanglement dynamics responsible for coherent oscillations while, as we will show below, it still accommodates qualitative features of chaotic behavior as the circuit parameters or initial conditions are varied.  

The set of all states generated by the circuit in Eq.~(\ref{eq:shallow_circuit}) spans a three-dimensional manifold $\mathcal{M}=\{\,|\psi(\bm z)\rangle\,|\,\bm z\in[0,2\pi)^3\}$ embedded within the exponentially large Hilbert space, see Fig.~\ref{fig:overview}{\bf b}.
We use TDVP~\cite{kramer1981geometry} to describe the time evolution within $\mathcal{M}$: the exact Schr\"odinger dynamics is projected onto the tangent space of $\mathcal{M}$, yielding effective classical equations of motion for the variational parameters $\bm z$. 
The equations of motion are nonlinear, reflecting the restricted structure of the manifold and its geometry, see Methods for their explicit form and the SM~\cite{SOM} for the full derivation. 

Within the TDVP framework, the projected dynamics in $\bm z$-space reveal qualitatively different behaviors depending on the initial state. As illustrated in Fig.~\ref{fig:overview}{\bf c}, some trajectories remain confined to quasi-periodic orbits, reminiscent of KAM tori, while others wander irregularly and spread throughout parameter space. The coexistence of these regular and irregular trajectories is a defining feature of mixed phase space. A powerful visualization tool of this coexistence is the Poincar\'e section~\cite{StrogatzBook}: by recording points each time a trajectory intersects a chosen plane, e.g., $\phi_1=0 \mod 2\pi$, the effective dynamics are reduced to a two-dimensional slice. As shown in Fig.~\ref{fig:overview}{\bf d}, regular trajectories appear as smooth closed curves corresponding to slices of invariant tori, since the orbit continues to intersect the plane in the vicinity of the original location even at late times. By contrast, chaotic trajectories manifest as scattered points filling extended regions, since chaotic dynamics take the system far away from the original intersection at later times. As we will show below, the location, size, and fine structure of these features depend sensitively on $(J_o,J_e,J_3)$, in analogy with the KAM theorem in classical mechanics. 

The shallow-circuit ansatz combined with TDVP projection provides a compact description of the interacting quantum model, Eq.~(\ref{eq:Hamiltonian}), displaying hallmarks of classical nonlinear systems, such as periodic trajectories embedded in a chaotic background. One may wonder whether the same features could be captured within the conventional semiclassical description based on  large-$S$ spin coherent states. This is not possible because the entanglement between neighboring spins is essential for describing their dynamics. On the other hand, the variational-manifold approach naturally represents a type of semiclassical limit: weakly-entangled ansatz form an overcomplete basis of the Hilbert space, and the TDVP equations correspond to the saddle-point trajectories of a path integral over these states~\cite{Haegeman2014geometry}. In this sense, the trajectories generated by the circuit in Fig.~\ref{fig:overview} are in direct correspondence with full quantum dynamics, as we demonstrate next. 

\begin{figure*}[t]
    \centering   \includegraphics[width=0.95\linewidth]{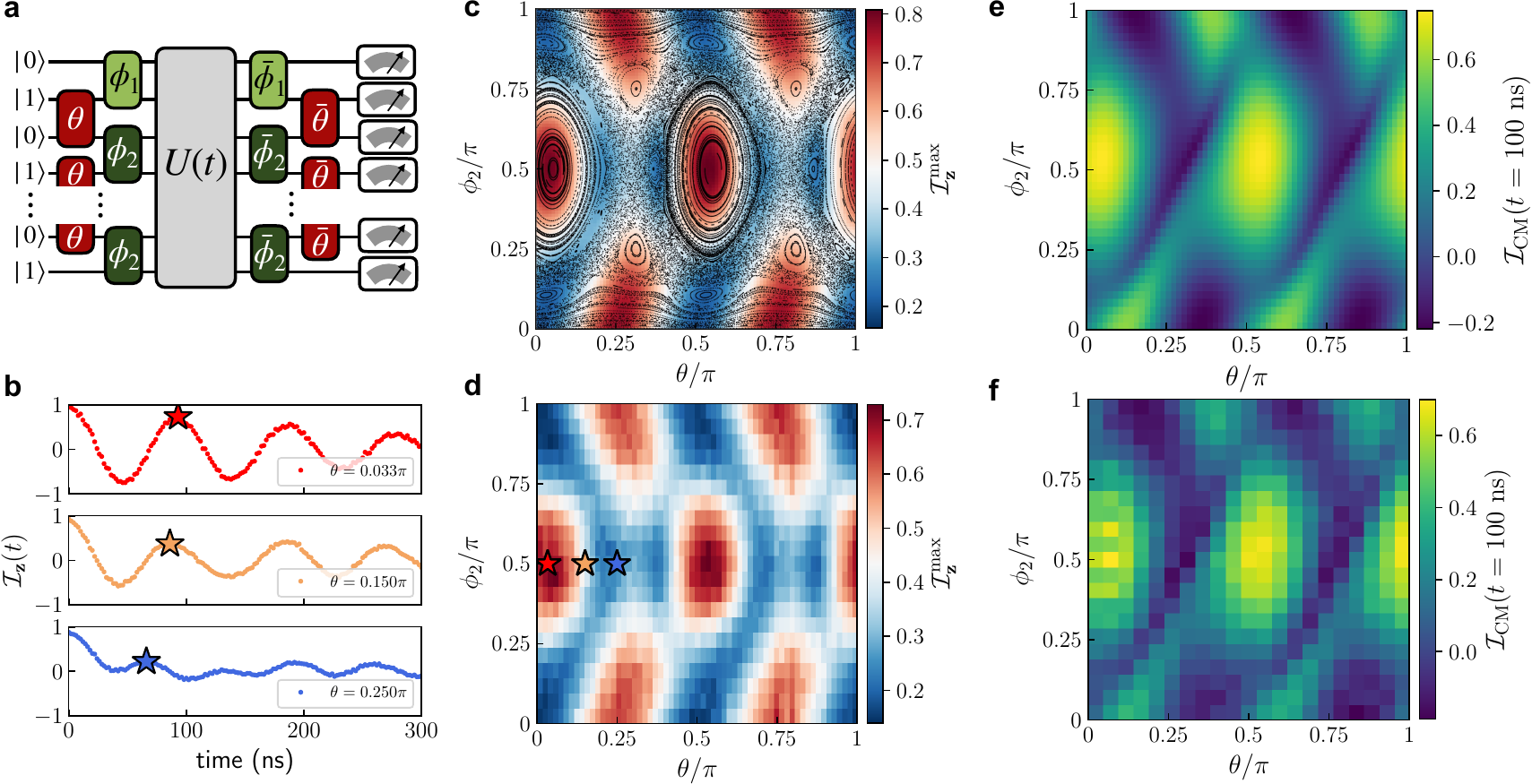}
    \caption{
    \textbf{Experimental mapping of the Poincar\'e section.} 
    \textbf{a} Quantum circuit implementing the measurement of imbalance $\mathcal{I}_{\bm z}(t)$ in Eq.~(\ref{eq:imbalance}). The system is initialized in a product state parameterized by $(\theta,\phi_1,\phi_2)$ by acting with a unitary $\hat U_\mathcal{M}$ in Eq.~(\ref{eq:shallow_circuit}), then evolved by the unitary $\hat U(t)$ generated by the Hamiltonian in Eq.~(\ref{eq:Hamiltonian}), and finally acted upon by the inverse circuit $\hat U_\mathcal{M}^\dagger$ before readout. 
    \textbf{b} Imbalance dynamics $\mathcal{I}_{\mathbf z}(t)$ for three representative initial states with fixed $\phi_2=0.5\pi$ and $\theta/\pi = 0.033$, $0.150$, and $0.250$ (red, orange, and blue, respectively). The star on each curve indicates the time at which the \emph{first revival peak} is identified; this defines the quantity $\mathcal{I}_{\bm z}^{\mathrm{max}}$ whose values are shown across phase space in panels~\textbf{c} and~\textbf{d}.   
    \textbf{c},~\textbf{d} First revival peak of the subsystem imbalance, $\mathcal{I}_{\bm z}^\mathrm{max}$, mapped across the phase space $(\theta,\phi_2)$ with $\phi_1 = 0$. 
    Panel \textbf{c} shows numerical results obtained from exact diagonalization of a 16-qubit chain with periodic boundary conditions, overlaid with the TDVP Poincar\'e section (black dots), illustrating the close correspondence between the two. 
    Panel \textbf{d} shows the corresponding experimental measurement on the quantum processor using 24 qubits. 
    The three initial conditions in~\textbf{b} are marked in~\textbf{d} by red, orange, and blue stars, indicating their locations within the regular island and the surrounding chaotic region. 
    The system is evolved for a sufficient amount of time (between $50$ and $120$~ns) to capture the first peak of $\mathcal{I}_{\bm z}(t)$.
    \textbf{e},~\textbf{f} Co-moving imbalance $\mathcal{I}_\mathrm{CM}$, representing the imbalance of the full quantum state along the classical trajectory predicted by TDVP, evaluated at a fixed time $t=100~\mathrm{ns}$. Panel {\bf e} is the numerical result, while panel {\bf f} shows the data from the quantum processor. The regular trajectories remain confined to the manifold, while chaotic trajectories experience pronounced leakage outside the manifold.
    In panels {\bf b}-{\bf f}, the imbalance was calculated for the 12 qubits in the middle of the chain.}
    \label{fig:poincare}
\end{figure*}

\section*{Mapping the Poincar\'e section}

While the TDVP framework allows us to visualize mixed phase space through Poincar\'e sections, such diagnostics are not directly accessible in experiment: constructing a Poincar\'e section requires tracking the effective phase-space variables $\bm z$, which are emergent coordinates of the variational manifold rather than physical observables of the qubits. Signatures of regular and chaotic behavior can, in principle, be revealed in the dynamics of local observables. However, due to the limited coherence time of the hardware, such quantities may not be sufficiently sensitive to reveal well-defined KAM tori as in Fig.~\ref{fig:overview}{\bf d}. To overcome these limitations, we introduce a closely related and experimentally accessible quantity based on imbalance of a subsystem $A$: 
\begin{equation}\label{eq:imbalance}
    \mathcal{I}_{\bm z} = 
    \frac{1}{N} \sum_{j\in A} (-1)^{j-1} 
    \langle\psi|\hat U_{\mathcal M}(\bm z)\hat\sigma^z_j 
    \hat U_{\mathcal M}^\dagger(\bm z)|\psi\rangle.
\end{equation}
The corresponding measurement protocol is summarized in
Fig.~\ref{fig:poincare}{\bf a}: the variational states $\ket{\psi(\bm z)}$ are prepared with the help of the unitary $\hat U_\mathcal{M}$ in Eq.~(\ref{eq:shallow_circuit}), evolved under the Hamiltonian (\ref{eq:Hamiltonian}) using the calibrated couplings and scheme in Fig.~\ref{fig:overview}, and subsequently reversed by an inverse unitary $\hat U_\mathcal{M}^\dagger$ before measuring the imbalance.

To gain some intuition about $\mathcal{I}_{\bm z}$, note that for the trivial circuit $\hat U_{\mathcal M}(\bm 0)= \hat{\mathbb{1}}$, $\mathcal{I}_{\bm 0}$ simply measures the difference between the occupations of odd and even sites, i.e., staggered magnetization.  For general parameters $\bm z$, $\mathcal{I}_{\bm z}$ quantifies the overlap between the quantum state $|\psi\rangle$ and the variational ansatz $|\psi(\bm z)\rangle$. If $|\psi\rangle$ is given by Eq.~(\ref{eq:shallow_circuit}), applying the inverse circuit $\hat U_{\mathcal M}^\dagger(\bm z)$ maps it back to the N\'eel state, maximizing the imbalance. Consequently, during the dynamics, $\mathcal{I}_{\bm z}(t)=1$ if and only if $|\langle\psi(0)|\psi(t)\rangle|^2=1$. Thus, $\mathcal{I}_{\bm z}$ serves as a practical proxy for global state fidelity, with the advantage of smaller measurement fluctuations compared to the latter (see Methods). Furthermore, $\mathcal{I}_{\bm z}$ is conveniently normalized: starting from an ansatz state $|\psi(\bm z)\rangle$, $\mathcal{I}_{\bm z}(t=0)$ is initially equal to unity and takes values strictly within the interval $[-1,1]$ at later times, providing a consistent scale for comparing different initial conditions. This normalization, together with its direct experimental accessibility, makes the imbalance an ideal observable for probing the underlying mixed structure of the dynamics. 

Figure~\ref{fig:poincare}{\bf b} compares the imbalance dynamics $\mathcal{I}_{\bm z}(t)$ for three representative initial states, with the subsystem $A$ being the central 12 qubits. When the system is initialized within a regular region of the TDVP phase space, $\mathcal{I}_{\bm z}(t)$ displays long-lived coherent oscillations with pronounced revivals characteristic of quasi-periodic motion. In contrast, an initial state chosen from the chaotic region exhibits a much faster decay of $\mathcal{I}_{\bm z}(t)$ with strongly suppressed revivals, signaling fast thermalization. We define the first-revival peak $\mathcal{I}_{\bm z}^\mathrm{max}$ as the value of $\mathcal{I}_{\bm z}(t)$ at its first local maximum, and use it as a measure of dynamical regularity for each initial condition. Scanning over $(\theta,\phi_2)$ yields a Poincar\'e-section-like map constructed entirely from local observables, i.e., without reconstructing the full variational coordinates $(\theta,\phi_1,\phi_2)$. 

Figures~\ref{fig:poincare}{\bf c}–{\bf d} present a direct comparison between numerical simulations and experimental measurements of the Poincar\'e section. In both cases, trajectories initialized within regular regions of the TDVP phase space in Fig.~\ref{fig:overview}{\bf c} display pronounced revivals of $\mathcal{I}_{\bm z}(t)$, consistent with their proximity to KAM-like islands. By contrast, trajectories launched in chaotic regions exhibit rapidly decaying imbalance, indicating fast thermalization and loss of coherence. This sharp dynamical contrast provides a direct quantum analog of the classical Poincar\'e section, with imbalance revivals faithfully reflecting the underlying phase space structure. The close correspondence between simulation and experiment highlights the robustness of imbalance as a scalable proxy for fidelity, while the global features of the resulting Poincar\'e-section-like diagram (Fig.~\ref{fig:overview}{\bf c}) align closely with TDVP predictions. As shown in the SM~\cite{SOM}, the structure of the Poincaré section is insensitive to the size of subsystem $A$ used to compute $\mathcal{I}_{\bm z}(t)$.

The variational manifold constitutes only an approximation to the full quantum dynamics and their discrepancy can be quantified using quantum leakage~\cite{wenwei18TDVPscar}. The latter is difficult to directly measure; instead, we consider the \emph{co-moving imbalance}, $\mathcal{I}_\mathrm{CM}\equiv \mathcal{I}_{\bm z(t)}$, where $\bm z(t)$ is the classical trajectory defined by the TDVP equations of motion. $\mathcal{I}_\mathrm{CM}$ evaluates the imbalance of the full quantum state along the TDVP-predicted path, and it should remain close to unity if TDVP faithfully reproduces the dynamics of experimentally accessible observables. The co-moving imbalance map at a fixed time slice is shown in Fig.~\ref{fig:poincare}{\bf e}-{\bf f}. Its phase-space profile confirms that the TDVP manifold accurately captures the coherent, low-entanglement sector of the evolution featuring KAM-like trajectories that remain confined to $\mathcal M$, while chaotic trajectories experience more pronounced leakage into higher-entanglement directions outside $\mathcal{M}$.

We note that there are small discrepancies between the Poincar\'e section in Fig.~\ref{fig:overview} and that inferred from experimental measurements, particularly near the upper and lower phase-space boundaries in Fig.~\ref{fig:poincare}{\bf c}-{\bf d}. These deviations are expected because decoherence effects restrict our measurements to much shorter times compared to the classical analysis in Fig.~\ref{fig:overview}; hence we can only resolve orbits with relatively short periods, i.e., the largest regular islands. In the SM~\cite{SOM}, we demonstrate that the agreement between TDVP and experiment can be improved by considering later-time imbalance dynamics. An additional source of discrepancy is the simple form of the shallow variational circuit, which may not capture finer details of quantum dynamics. While this could be systematically improved by refining the ansatz, our minimal parametrization in Eq.~(\ref{eq:shallow_circuit}) is sufficient to capture the essential aspects of the global mixed-phase structure.

\begin{figure*}[t]
    \centering   \includegraphics[width=0.9\linewidth]{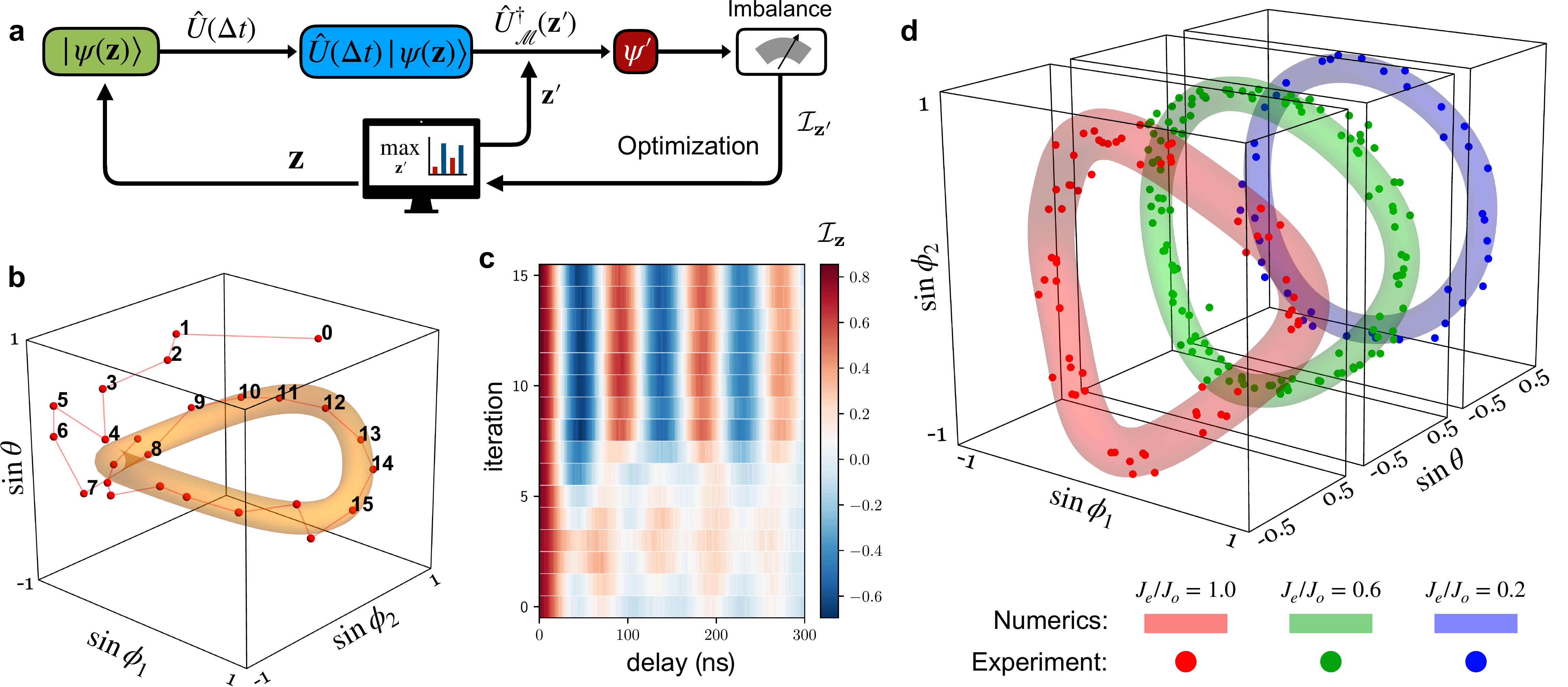}
    \caption{
    \textbf{Hybrid feedback control of variational dynamics.} 
    \textbf{a} Schematic workflow of the iterative feedback loop in Eq.~(\ref{eq:iterative}). Starting from an initial state $|\psi(\bm z)\rangle$ with parameters $(\theta,\phi_1,\phi_2)$, the system evolves under the unitary $\hat U(\Delta t)$ generated by the Hamiltonian in Eq.~(\ref{eq:Hamiltonian}). An inverse variational circuit $\hat U^{\dagger}_{\mathcal M}(\bm z')$ is then applied, and the measured imbalance $\mathcal{I}_{\bm z'}$ is maximized through classical optimization, providing updated parameters $\bm z'$ for the next iteration.
    \textbf{b} Feedback evolution for a single experimental trajectory on the superconducting processor using 24 qubits, using a time interval $\Delta t = 80~\mathrm{ns}$ for each feedback step. Starting from $\bm{z}^{(0)}=(\pi/4,0,\pi/2)$, the markers indicate the first few iteration steps, showing the convergence of the parameters toward a closed periodic orbit (orange surface) predicted by numerical simulation. 
    \textbf{c} Dynamics of imbalance $\mathcal{I}_{\bm z}$ for the 12 central sites for each of the iteration steps marked in panel {\bf b}, illustrating the steady convergence to the periodic orbit.
    \textbf{d} The stability of mixed phase space as the even-bond coupling $J_e$ is varied, with fixed $J_o/2\pi=5.0~\mathrm{MHz}$ and $J_3/2\pi=0.25~\mathrm{MHz}$.
    Different plots correspond to $J_e/2\pi=5.0~\mathrm{MHz}$ (front, red), $J_e/2\pi=3.0~\mathrm{MHz}$ (middle, green), and $J_e/2\pi=1.0~\mathrm{MHz}$ (back, blue). In each case, the hybrid protocol converged to a stable orbit, whose geometry and curvature vary continuously with $J_e$, reflecting the smooth deformation of the mixed phase-space structure. Data points are experimental results, while the shaded surfaces represent the corresponding stable orbits obtained by numerical implementation of the protocol. The width of each shaded surface reflects the standard deviation of the experimental data from the numerical prediction.}
    \label{fig:scarfinder}
\end{figure*}

\section*{Hybrid feedback control}

Characterizing mixed phase-space structures in high-dimensional manifolds is challenging: the Poincar\'e section captures only a low-dimensional slice of the dynamics, rapidly becoming impractical to analyze and losing its visual intuition as the dimension of $\mathcal{M}$ grows. To overcome this, we develop a hybrid feedback protocol that combines short quantum evolutions with classical optimization, which allows to directly ``filter out'' nonthermal trajectories on quantum hardware. Beyond finding such trajectories, this will allow us to verify the robustness of the mixed phase-space structure, as we expect the regular islands to deform smoothly upon varying the system's parameters. In classical mechanics, such structural stability is a defining feature of mixed phase spaces: weak perturbations modify but do not destroy the coexistence of order and chaos.

To steer quantum dynamics towards a regular region of phase space, we employ a hybrid quantum--classical optimization scheme inspired by the \textsc{ScarFinder} algorithm~\cite{scarfinder}. The key idea is that coherent trajectories remain confined within $\mathcal{M}$, whereas thermalizing components tend to rapidly escape from it. Thus, short-time evolution $\hat U(\Delta t) = e^{-i\hat H\Delta t/\hbar}$ slightly drives the state away from $\mathcal{M}$, while projection back onto the manifold, $\mathcal{P}_\mathcal{M}$, removes the entanglement-generating component and restores coherence. Repeated application of this evolution--projection cycle acts as a dynamical filter:
\begin{equation}\label{eq:iterative}
    |\psi^{(n+1)}\rangle = 
    \mathcal{P}_{\mathcal{M}}\, \hat U(\Delta t)\,|\psi^{(n)}\rangle,
\end{equation}
where $\mathcal{P}_{\mathcal{M}}$ is implemented through feedback optimization. Under iteration, 
regular motion governed by TDVP flow is reinforced, converging toward a self-consistent orbit that minimizes leakage from $\mathcal{M}$.

The experimental workflow of the hybrid feedback procedure is illustrated in Fig.~\ref{fig:scarfinder}{\bf a}, see Methods for details. Starting from an initial state $|\psi(\bm z^{(0)})\rangle$, the system is evolved under the Hamiltonian in Eq.~(\ref{eq:Hamiltonian}) for a short time $\Delta t$. A reverse variational circuit $\hat U_{\mathcal M}^\dagger(\bm z')$ with adjustable parameters $\bm z'$ is then applied to this state, followed by local measurement of the imbalance $\mathcal{I}_{\bm z'}$. The measured imbalance is processed by a classical optimizer, which updates $\bm z'$ to maximize $\mathcal{I}_{\bm z'}$, producing a refined parameter set $\bm z^{(1)}$ that defines the projected state $|\psi(\bm z^{(1)})\rangle$ on the variational manifold $\mathcal{M}$. This new state is re-prepared on the processor to initiate the next iteration. Repeating this sequence (short-time evolution, inverse circuit, measurement, and feedback) generates a discrete trajectory $\{\bm z^{(n)}\}$ in the variational space. Convergence is achieved when the parameters stabilize, indicating that the feedback loop (\ref{eq:iterative}) has locked onto a self-consistent periodic orbit of the hybrid map.

Intuitively, our hybrid loop operates as a closed learning process that ``distills'' coherence from chaotic motion: each forward evolution step introduces small deviations due to entanglement and experimental noise, while each feedback projection re-aligns the state with the manifold. The resulting steady orbit corresponds to the most regular trajectory accessible within $\mathcal{M}$ -- a dynamically stabilized mode that simultaneously suppresses entropy growth and enhances revival fidelity. This self-correcting feedback mechanism makes the hybrid scheme particularly suitable for noisy quantum devices, where short, accurate operations can be repeated even though long coherent evolution is not achievable. It can be straightforwardly extended to deeper variational circuits by optimizing over additional layers or parameters within the same iterative framework.

We implemented the hybrid quantum--classical feedback protocol on the superconducting processor using 24 qubits, see Fig.~\ref{fig:scarfinder}{\bf b}.
We initialize a representative state with parameters $(\theta,\phi_1,\phi_2)=(\pi/4,0,\pi/2)$, residing deep within the chaotic region of the TDVP phase space. After fewer than 10 iterations of the hybrid protocol, this trajectory converges to the optimal classical orbit predicted by TDVP (orange surface). We note that the shape of this orbit depends on $\Delta t$: finite $\Delta t$ introduces a small deformation of the classical trajectory, and this deviation vanishes in the limit $\Delta t \!\to\! 0$~\cite{SOM}. This deformation is intentionally leveraged in the experiment: we adopt $\Delta t = 80$ ns, which is close to the system's intrinsic timescale, $\hbar/J_o \approx 200$ ns, in order to ensure rapid convergence of the hybrid feedback loop. 
The dynamics of imbalance $\mathcal{I}_{\bm z}(t)$ at each iteration step, measured on the processor using the same method as Fig.~\ref{fig:poincare}{\bf b}, are shown in Fig.~\ref{fig:scarfinder}{\bf c}. These demonstrate that the feedback algorithm reliably converges to a stable trajectory, exhibiting persistent oscillations in $\mathcal{I}_{\bm z}(t)$, even in the presence of device noise, crosstalk, and calibration imperfections. The discrete parameter trajectories $\{\bm z^{(n)}\}$, extracted from successive feedback iterations, gradually approach the periodic orbit, tracing its basin of attraction in parameter space. Small fluctuations around the converged trajectory seen in Fig.~\ref{fig:scarfinder}{\bf c} are attributed to residual experimental noise rather than instability of the algorithm~\cite{SOM}. 

Finally, in Fig.~\ref{fig:scarfinder}{\bf d} we directly probe the stability of mixed phase space by varying the coupling strength. Fixing $J_o/2\pi=5.0~\mathrm{MHz}$ and $J_3/2\pi=0.25~\mathrm{MHz}$, we run the previously described hybrid protocol for different values of $J_e/2\pi=1.0,3.0$, and $5.0~\mathrm{MHz}$. In each case, the feedback dynamics identify a family of stable periodic orbits whose curvature and extent systematically evolve with $J_e$. This continuous deformation of the orbits reflects the reshaping of the mixed phase space structure and confirms that the geometry of regular regions can be directly tuned by the underlying Hamiltonian parameters.  We emphasize that the same iterative method also provides a means of preparing long-lived nonthermal states, without requiring any input about the model beyond the definition of the manifold $\mathcal{M}$.

\section*{Conclusions and outlook}

By combining analog quantum evolution with classical feedback optimization, we demonstrated a hybrid control protocol that uncovers and stabilizes long-lived coherent trajectories in quantum many-body systems. Each feedback cycle requires only short coherent evolutions followed by local measurements and classical optimization, inherently minimizing decoherence and calibration errors and avoiding the need to reconstruct the full many-body wave function. As a proof-of-principle, we revealed a regime where regular and chaotic trajectories coexist within the same interacting qubit model, providing the first experimental evidence for a mixed phase space in a quantum many-body system. 

The mixed phase space observed here bears some similarity with the phenomenon of quantum many-body scars, as the latter are also associated with regular dynamics albeit from a few special initial states~\cite{Bernien2017,Turner2017}. By analogy with single-particle quantum scars~\cite{Heller84}, one expects the regular dynamics to stem from \emph{isolated} unstable periodic orbits of the underlying classical system. In contrast, our feedback-based probe reveals that scarred dynamics are seamlessly embedded within a larger mixed-phase-space structure of coexisting regular and chaotic trajectories in Fig.~\ref{fig:poincare}. Indeed, as shown in the SM~\cite{SOM}, the reduced TDVP equations for a special choice of couplings and initial conditions reproduce the oscillatory trajectories of the ``rainbow'' quantum many-body scars~\cite{Langlett2021,Dong2023}. This underscores the need for a better understanding of the origin of quantum many-body scarring in relation to its single-particle counterpart~\cite{wenwei18TDVPscar,Turner2020,Evrard2024,Pizzi2025} and the phenomenon of mixed phase space reported here. 

More broadly, our protocol can be viewed as a generalization of classical chaos control~\cite{Ott1990,Antoniou1997}, where unstable periodic orbits are stabilized through continuous feedback, to the quantum domain. The core idea of utilizing feedback to amplify ergodicity breaking has intriguing connections with adaptive quantum dynamics~\cite{Sierant2023,Roy2020,Herasymenko2023,ODea2024} and phase transitions induced by measurements~\cite{Li2018,Vasseur2019,Nahum2021,Ippoliti2021} and control~\cite{Iadecola2023,Iadecola2025}. 
Moreover, our protocol complements existing optimization strategies for quantum state preparation based on matrix-product-state methods~\cite{Doria2011,Jensen2021,Ljubotina2022Steering}, machine learning~\cite{Bukov2018,Metz2023}, and adiabatic evolution~\cite{delCampo2013,Sels2017}.  
All of these approaches can be naturally blended with our feedback framework to explore how semiclassical behavior emerges and can be controlled in broader classes of many-body Hamiltonians, with straightforward extensions to deeper variational circuits and higher dimensions.

\section*{Acknowledgments}
 We acknowledge Haohua Wang for supporting facilities that cover full experiments of this work from device fabrication to measurement.  The device was fabricated at the Micro-Nano Fabrication Center of Zhejiang University. The authors at Zhejiang University acknowledge the support of the National Key R\&D Program of China (Grant No.2024YFA1408900 and No.2022YFA1404203), the Pioneer R\&D Program of Zhejiang (No.2025C01046 and No.2025C01019), the National Natural Science Foundation of China (Grant No.12504590, No.92365301, No.12322414, No.12274367, No.12375021, No.12404570, No.12274368), the Zhejiang Provincial Natural Science Foundation of China (Grant No.LDQ23A040001 and No.LR24A040002). J.R., A.H., and Z.P. acknowledge support by the Leverhulme Trust Research Leadership Award RL-2019-015 and EPSRC Grants EP/Z533634/1, UKRI1337. This research was supported in part by grant NSF PHY-2309135 to the Kavli Institute for Theoretical Physics (KITP). 

\vspace*{0.2cm}

\section*{Author Contributions Statement}

J.R., A.H., L.Y. and Z.P. proposed the model and algorithmic protocol, and conducted the theoretical analysis. H.D. carried out the experiment and analysed the data under the supervision of Z.W.  H.D., J.R. and H.W. performed the numerical simulation. H.L., Q.G., Z.C., and Y.Z. fabricated the device. J.R., H.D., A.H., L.Y. and Z.P. co-wrote the manuscript with input from other authors. All authors participated in the discussions of the results and the development of the manuscript.

\section*{Competing Interests Statement}

The authors declare no competing interests.

\bibliography{bibliography}


\section*{Methods}

\subsection*{Experimental setup}
Our experiments are conducted on a 2D flip-chip superconducting quantum processor comprising 125 frequency-tunable transmon qubits, with each adjacent pair coupled via a tunable coupler. A ladder configuration consisting of 24 qubits and 34 couplers (Fig.~\ref{fig:overview}{\bf a}) is employed with precisely calibrated coupling strengths of approximately 0-15 MHz to realize the target Hamiltonian in Eq.~(\ref{eq:Hamiltonian}), see SM~\cite{SOM} for detailed characterization of the device.

To observe the Poincar\'e section and implement hybrid feedback control, our experimental protocol integrates both digital quantum circuits and analog quantum simulation. The procedure starts with the execution of shallow quantum circuits, after which all qubits are switched from their individual circuit's frequency to the resonance frequency at 3.7 GHz. When qubits are brought into resonance at the operating point, the inter-qubit couplings are dynamically activated to implement the desired Hamiltonian evolution with interaction strengths controlled via tunable couplers. Following evolution periods of the desired time, the qubits are returned to their idle frequencies to execute the reversed gate sequence.
Throughout these reference frame transformations, all qubits accumulate additional phases that are compensated in our calibration protocol~\cite{SOM}.

The processor features capacitively coupled readout resonators designed at around 6.4 GHz for dispersive measurement, enabling simultaneous qubit state detection with average fidelity exceeding $98\%$.
The system is integrated using wire-bonding techniques and protected by multilayer magnetic shields in a dilution refrigerator operating at 15 mK.

\subsection*{Shallow circuit implementation}

The shallow circuit is implemented by compiling the required XY gates as the combination of single-qubit gates and CZ-gates as follows:
\begin{equation*}
\resizebox{\linewidth}{!}{
$\text{XY}(\theta) = 
\begin{quantikz}[column sep=0.03cm,row sep=0.03cm]
\qw &\gate{R_x(\pi/2)} &\qw& \ctrl{1} & \gate{R_z(-\pi/2)}  &\gate{R_y(\theta+\pi)}  & \ctrl{1} &\gate{R_y(-\pi/2)}&\gate{R_z(\pi/2)}& \qw \\
\qw &\gate{R_y(\pi/2)} & \gate{R_z(\pi)} & \ctrl{-1} & \gate{R_y(-\theta)}& \qw & \ctrl{-1} &\gate{R_y(-\pi/2)}& \qw& \qw 
\end{quantikz}$}
\end{equation*} 
Single-qubit gates are implemented using 20 ns microwave pulses with a Gaussian envelope, optimized via the derivative reduction by adiabatic gate (DRAG) technique. The  CZ gates are realized by tuning $|11\rangle$ and $|20\rangle$ states of the two qubits near resonance and switching 
on the coupling by applying a fast flux pulse to the corresponding coupler between them for a duration of 32 ns. Benchmarking with simultaneous cross-entropy benchmarking protocols reveals single-qubit gate fidelities exceeding 99.95\% and two-qubit CZ gate fidelities above 99.5\%.
Additionally, arbitrary single-qubit rotations are decomposed into a virtual Z-phase gate followed by an actual XY rotation. The details of the fidelity of our gates are shown in the SM~\cite{SOM}.

\subsection*{Equations of motion for the shallow circuit ansatz}

By projecting quantum dynamics generated by the Hamiltonian in Eq.~(\ref{eq:Hamiltonian}) onto the variational manifold spanned by states in Eq.~(\ref{eq:shallow_circuit}), we obtain the following TDVP equations of motion:
\begin{equation}
\begin{aligned}\label{eq:tdvpeom}
    \dot \theta &= J_e \cos\phi_1\cos\phi_2 - J_3 \cos^2(2\theta)\, \sin\phi_1\sin\phi_2, \\
    \dot \phi_1 &= J_o
	+ \frac{\sin(4\theta)\,\big(J_e \sin\phi_1 \cos\phi_2 + J_3 \sin\phi_2 \cos\phi_1\big)}{1+\cos^2(2\theta)}, \\
	\dot \phi_2 &= J_o
	+ \frac{\sin(4\theta)\,\big(J_e \sin\phi_2 \cos\phi_1 + J_3 \sin\phi_1 \cos\phi_2\big)}{1+\cos^2(2\theta)},
\end{aligned}
\end{equation}
see the SM~\cite{SOM} for the full derivation. By integrating these equations, we obtain the phase-space portrait in Fig.~\ref{fig:overview}{\bf d} of the main text. In contrast to the underlying linear Schr\"{o}dinger equation, the projected dynamics generated by Eqns.~(\ref{eq:tdvpeom}) are evidently nonlinear. On the one hand, this nonlinearity allows for the emergence of mixed phase space, while on the other, it implies that the TDVP evolution cannot perfectly reproduce the exact quantum dynamics. As we enlarge the variational manifold, we can represent increasingly finer details of the full quantum dynamics. With this caveat in mind, our variational ansatz is the minimal description that captures the essential dynamical features of the SSH ladder in an analytically tractable way, illustrating the power of this approach.  

\begin{figure}[tb]
    \centering   \includegraphics[width=0.98\linewidth]{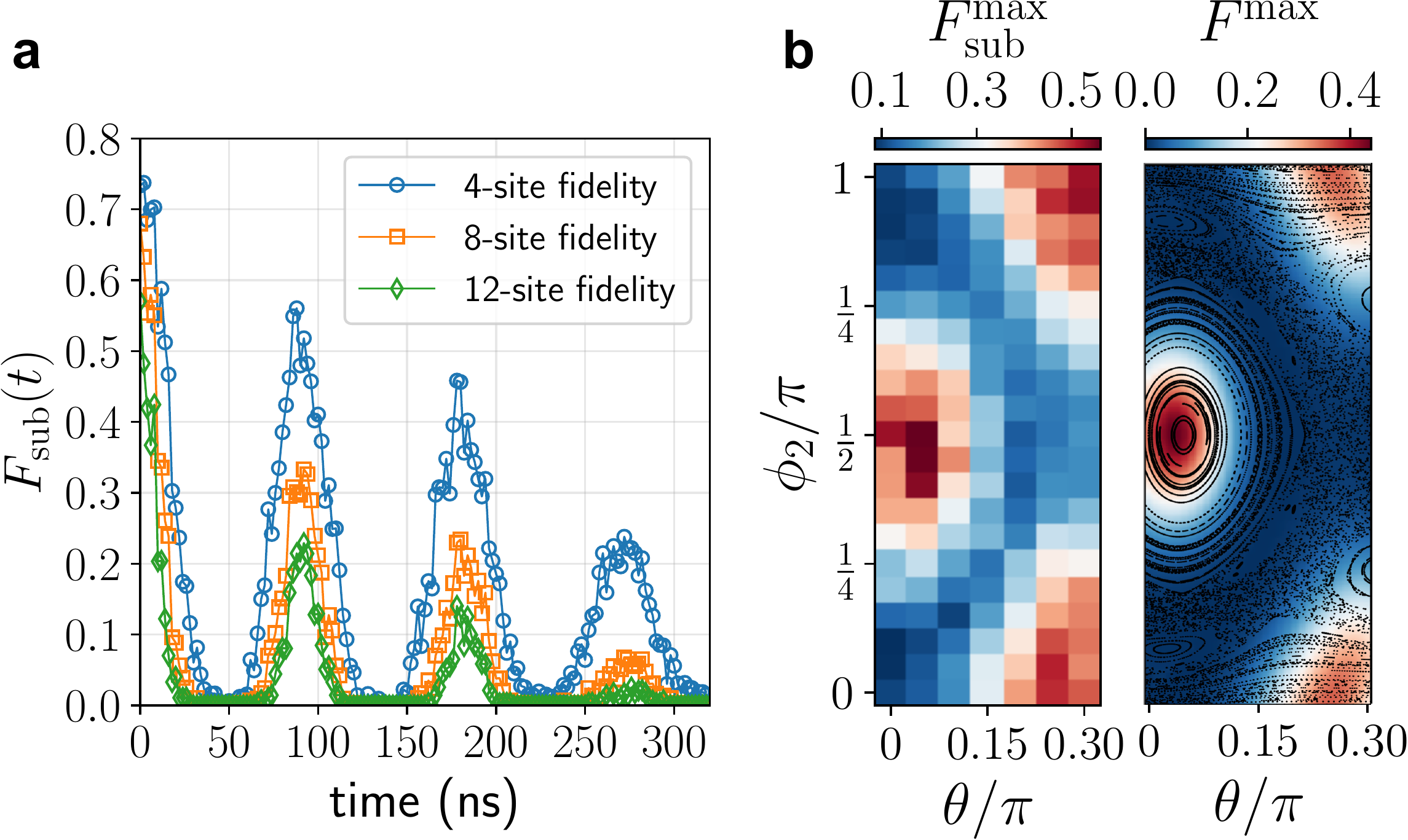}
    \caption{
    \textbf{Subsystem fidelity.}
    \textbf{a} Subsystem fidelity $F_{\rm sub}(t)$ for three choices of subsystem size $L_{\rm sub}=4,8,12$ and fixed initial condition $(\theta,\phi_2)=(0.05\pi,0.5\pi)$.  In all the measurements, the subsystem is taken as the central block of $L_{\rm sub}$ qubits.
    \textbf{b} Poincar\'e-section--like maps constructed from the first fidelity revival. The left panel shows the 4-qubit subsystem fidelity $F_{\rm sub}^{\max}(\theta,\phi_2)$ obtained experimentally. The right panel shows the corresponding \emph{global} fidelity $F^{\max}(\theta,\phi_2)$ obtained from exact diagonalization of an $L=16$ spin chain with periodic boundary conditions, overlaid with the TDVP Poincar\'e section (black dots).  
    Despite the larger fluctuations in the fidelity-based maps, the underlying phase-space structure, including the regular islands and chaotic regions, is visible in both cases. Larger subsystems more closely approximate the global fidelity but also exhibit increased noise.    
    }
    \label{fig:method_fid}
\end{figure}

\subsection*{Experimental measurement of fidelity}

Fidelity provides a natural measure of how closely a quantum system returns to its initial state under time evolution,
\begin{equation}
    F(t) = |\langle \psi(0) | \psi(t) \rangle|^2.
\end{equation}
In general, extracting $F(t)$ in a many-body system requires full quantum state tomography, which is exponentially costly in terms of the system size and therefore impractical for our quantum processor.

In our setting, however, the fidelity can be accessed more efficiently using a Loschmidt-echo–type protocol. The initial state $|\psi(0)\rangle$ is prepared by a known unitary circuit $\hat U_{\mathcal M}$ acting on a product state $\ket{0101\ldots}$. After evolving the system for a time $t$ under $\hat U(t)$, we apply the inverse circuit $\hat U_{\mathcal M}^{\dagger}$. This maps the overlap onto a measurement probability:
\begin{equation}
\begin{aligned}
F(t) = \Pr\!\Big[\, 
    &\text{measure} \ket{0101\ldots} \text{after } \hat U_{\mathcal M}^{\dagger}\hat U(t)
\Big],
\end{aligned}
\end{equation}
which counts how often the reference computational basis state is observed in bitstring snapshots.

However, for a 24-qubit system, the reference bitstring typically appears with exponentially small probability, far below the experimental noise floor. To obtain a robust signal, we therefore introduce the \emph{subsystem fidelity}, defined by checking whether a chosen block of $L_{\rm sub}$ consecutive qubits matches the reference pattern. For each snapshot, the subsystem fidelity contribution is one if the block matches and zero otherwise; averaging over all shots yields
\begin{equation}
\begin{aligned}
F_{\rm sub}(t) = \Pr\!\Big[ \text{measure}\, |\ldots \underbrace{0101\ldots01}_{L_\mathrm{sub}} \ldots\rangle \,
     \text{at time} \, t \Big].
\end{aligned}
\end{equation}
Because only $L_{\rm sub}$ spins are compared, $F_{\rm sub}(t)$ remains experimentally accessible even for moderate subsystem sizes.

Figure~\ref{fig:method_fid}{\bf a} shows $F_{\rm sub}(t)$ for several subsystem sizes, extracted for the initial condition $(\theta,\phi_2)=(0.05\pi,0.5\pi)$. Smaller subsystems produce smoother, less noisy behavior, while larger subsystems more closely approximate the global fidelity but exhibit stronger statistical fluctuations. For all choices of $L_{\rm sub}$, the revival structure follows that of the imbalance $\mathcal{I}_{\bm z}(t)$, demonstrating that both observables encode essentially the same dynamical information.

To visualize the dependence on initial conditions, we construct a Poincar\'e-section–like map using the first revival peak $F_{\rm sub}^{\max}$ for each $(\theta,\phi_2)$ in Fig.~\ref{fig:method_fid}{\bf b}. Despite increased noise and reduced contrast compared to the imbalance-based map $\mathcal{I}_{\bm z}^{\max}$ used in the main text, the qualitative phase-space structure is reproduced.

In summary, imbalance and fidelity diagnose regular versus chaotic dynamics in the same qualitative manner. The imbalance, however, is significantly simpler and yields higher-contrast maps across the full phase space. For this reason, we rely on the imbalance as the primary experimental observable in the main text, while fidelity-based measurements provide supporting validation of the results.

\subsection*{Algorithmic structure and physical interpretation of the hybrid protocol}

Our hybrid optimization protocol is implemented as an iterative feedback loop that alternates between short quantum evolutions and classical parameter updates. Each iteration approximates a discrete TDVP step, with the projection onto the variational manifold realized experimentally through measurement and optimization rather than analytically. 

We begin by preparing a state $|\psi(\bm z^{(0)})\rangle$ with parameters $\bm z^{(0)}=(\theta^{(0)},\phi_1^{(0)},\phi_2^{(0)})$ [Eq.~(\ref{eq:shallow_circuit})] on a quantum processor. We then iterate through the following sequence:
\begin{enumerate}
    \item {Time evolution:}  
    Evolve for a short interval $\Delta t$, 
    $|\psi\rangle = \hat U (\Delta t)|\psi(\bm z^{(0)})\rangle$, with the unitary $\hat U(\Delta t) = e^{-i\hat H \Delta t/\hbar}$ corresponding to the Hamiltonian $\hat H$.

    \item {Reverse circuit:}  
    Apply the inverse variational circuit $\hat U_{\mathcal M}^{\dagger}(\bm z')$ [Eq.~(\ref{eq:shallow_circuit})], parameterized by a trial set of variables $\bm z'$, to obtain
    $|\psi'\rangle = \hat U_{\mathcal M}^{\dagger}(\bm z')|\psi\rangle$.

    \item {Measurement:}  
    Measure the local spin expectations $\langle\hat\sigma^z_j\rangle$ and compute the imbalance $\mathcal{I}_{\bm z'}$ [Eq.~(\ref{eq:imbalance})]. 

    \item {Classical optimization:}  
    The measured imbalance is processed by a classical optimizer, which updates the parameters $\bm z'$ to maximize the overlap with the evolved state. The optimal parameters, 
    $\bm z^{(1)} = \arg\max_{\bm z'}\,\mathcal{I}_{\bm z'}[|\psi\rangle]$,
    define the best variational approximation $|\psi(\bm z^{(1)})\rangle$ to the evolved state $|\psi\rangle$.

    \item {Reinitialization:}  
    Prepare the updated state $|\psi\rangle=|\psi(\bm z^{(1)})\rangle$ and repeat the loop.
\end{enumerate}

 Experimentally, each function evaluation in step 4 corresponds to a complete quantum measurement. Typical convergence for the next evolved state requires $50$–$200$ iterations, with total experimental duration spanning approximately $0.5$-$2$ hours.
To mitigate finite-size boundary effects, we employ the imbalance (measured at $\Delta t=80$ ns) of the four central qubits as the observable for optimizing the parameters ${\bm z}^{(n)}$. The resultant optimal set is then used to reinitialize and reprepare the quantum state for subsequent measurement cycles.
Iterating this feedback cycle generates a discrete trajectory $\{\bm z^{(n)}\}$ within the variational manifold $\mathcal{M}$. Convergence is reached when the parameters stabilize along a closed orbit (typically within a few hundreds of iterations), indicating that the feedback loop has identified a self-consistent periodic trajectory under evolution and projection, which is the most stable dynamical mode supported by the manifold.

\clearpage
\onecolumngrid
\setcounter{page}{1}
\beginsupplement

\begin{center}
{\bf \large Supplemental Material for ``Quantum many-body mixed phase space revealed by hybrid feedback control''}    
\end{center}

\begin{center}
{\small This Supplemental Material contains additional information about our superconducting quantum processor and measurement procedure, a complete derivation of the equations of motion for the shallow circuit using time-dependent variational principle (TDVP), and further analyses of quantum many-body mixed phase space and the hybrid feedback protocol.}
\end{center}

\FloatBarrier

\section{Device performance}
\begin{figure}
    \centering
    \includegraphics[width=0.9\linewidth]{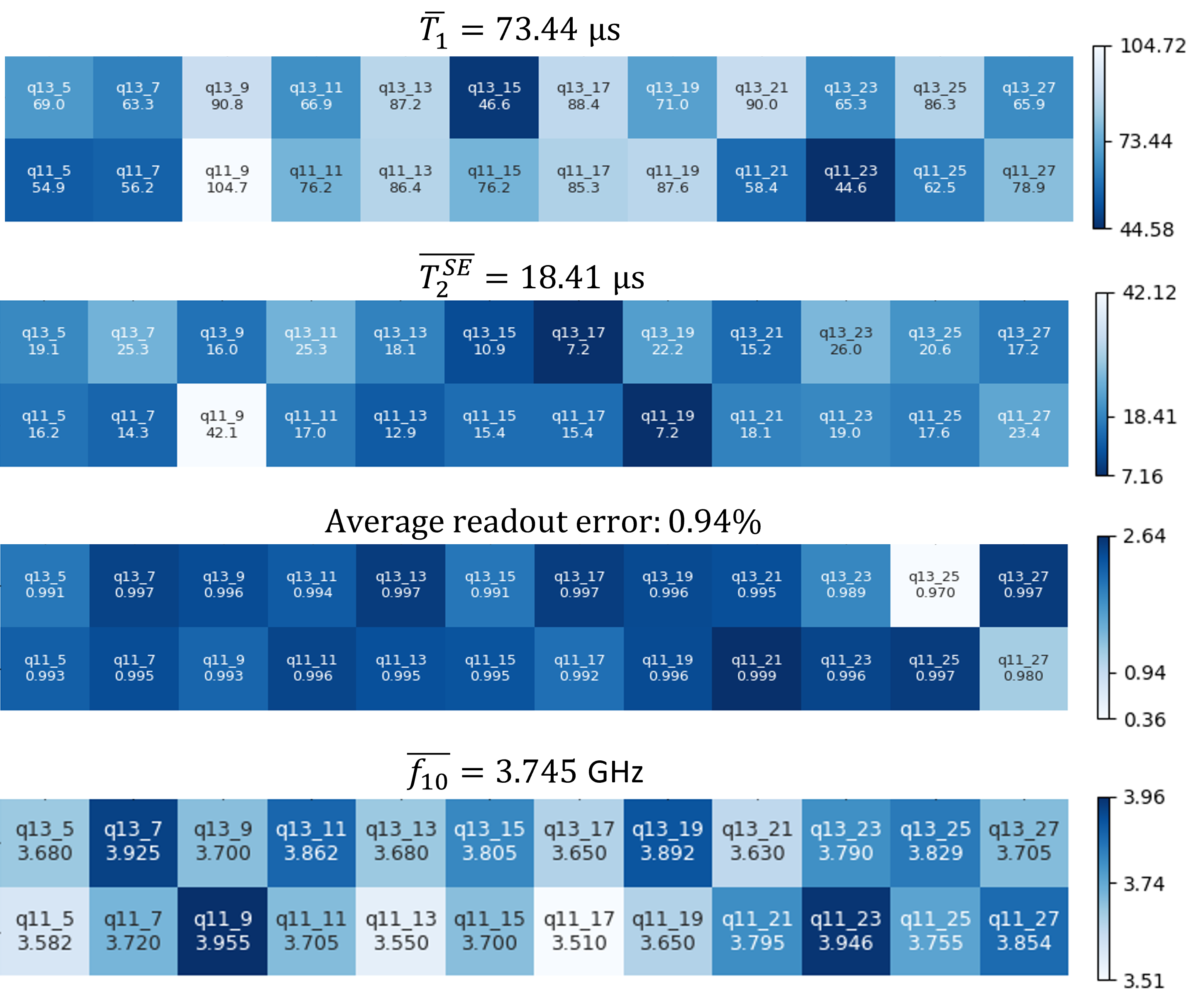}
    \caption{{\bf Heat map of single-qubit parameters.} Top to bottom:  Qubit relaxation time measured at its idle frequency, with  73.62 $\mu s$ the median for all qubits; Qubit dephasing time measured using Hahn echo sequence at its idle frequency, with 17.4 $\mu s$ the median for 24 qubits; Readout error averaged for qubit in $|0\rangle$ and $|1\rangle$, measured by preparing random product states on all qubits, with the median error rate 0.76$\%$; Qubit idle frequency.}
    \label{fig:deviceParameter}
\end{figure}

\begin{figure}
    \centering
    \includegraphics[width=0.99\linewidth]{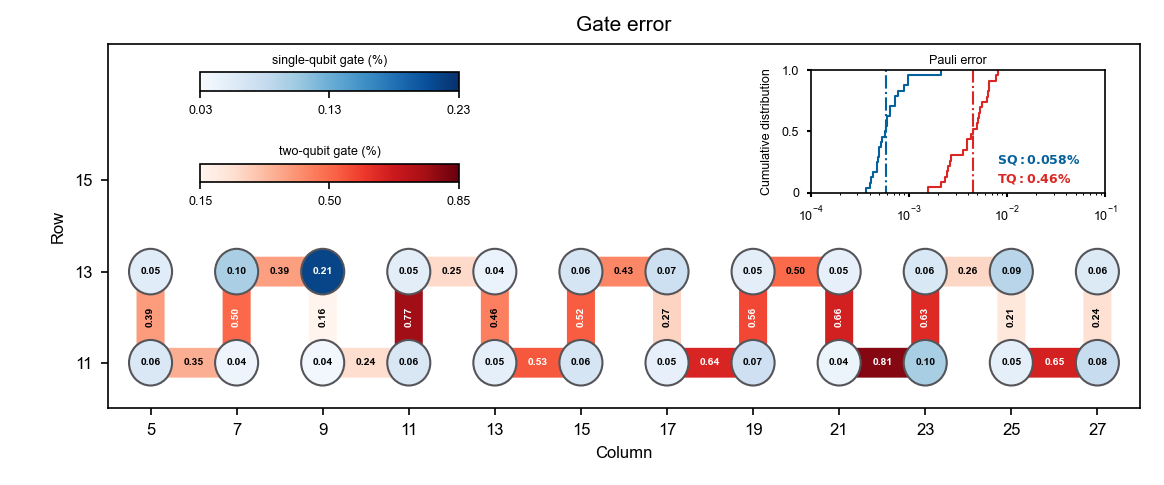}
    \caption{\textbf{Pauli errors of single-qubit and two-qubit gates.} Gate errors are benchmarked with simultaneous cross-entropy
    benchmarking (XEB). Errors of single-qubit gates (blue circles) are obtained by running single-qubit XEB sequences for all qubits
    simultaneously, while errors of two-qubit gates (red bars) are averaged over the two-qubit layers used
    in our experiments. For each two-qubit layer, we run two-qubit XEB sequences simultaneously for all the two-qubit gates in this layer. The inset shows the cumulative distribution of gate errors, with the dashed lines indicating the median values.}
    \label{fig:pauli_error}
\end{figure}

The wiring information and room temperature control electronics are similar to those of ~\cite{feitong2025prethermalZeroModes}. Figure~\ref{fig:deviceParameter} summarizes the typical single-qubit parameters of our processor, where single-qubit gates are applied at idle frequencies ranging from 3.5 GHz to 3.96 GHz. The energy relaxation time $T_1$ and dephasing time $T_2$ measured at these idle frequencies are also reported; both are substantially longer than the execution time of our shallow circuits. To enhance readout fidelity, an additional microwave pulse is applied to each qubit prior to the readout pulse, driving the $|1\rangle\leftrightarrow|2\rangle$ transition. The readout error, defined as the average misassignment probability for the $0\rangle$ and $1\rangle$ states, is predominantly below 1$\%$ across all qubits. All data presented in this work have been processed with readout error mitigation. 

As detailed in the Methods section, our experimental sequence involves compiling shallow circuits into combinations of single-qubit gates and two-qubit CZ gates to prepare the targeted states. During this process, all qubits are maintained at their idle frequencies, while the couplers are kept in a nearly off state. To minimize spurious interactions, we actively control not only the qubits forming the ladder geometry but also two adjacent rows of surrounding qubits and all associated couplers. Consecutive single-qubit gates are consolidated into a single XY microwave rotation combined with a virtual Z phase adjustment, thereby reducing the total number of physical gate operations. We characterize gate performance via simultaneous cross-entropy benchmarking (XEB), achieving median fidelities of approximately 99.95$\%$ for single-qubit gates and 99.5$\%$ for two-qubit gates, respectively (Fig.~\ref{fig:pauli_error}).

\section{Hybrid quantum simulation}

Our experimental protocol integrates digital quantum circuits with analog quantum simulation. The procedure begins with the execution of shallow quantum circuits, after which all qubits are switched from the gate-operation reference frame to the Hamiltonian evolution reference frame. When the qubits are tuned into resonance, the inter-qubit couplings are activated to realize the target Hamiltonian evolution. After evolution, the qubits are shifted from the resonance frequency of 3.7 GHz back to their respective idle frequencies to implement the reversed quantum gate sequence. Throughout this reference frame transformation, all qubits accumulate additional phases.

To calibrate and compensate for these phase shifts via single-qubit gates, we implement the following calibration procedure:

1. We first coarsely adjust all coupling strengths to match the target Hamiltonian parameters. To calibrate the resonance condition, we perform swap spectroscopy between each pair of qubits while detuning all other qubits by at least 100 MHz from the interaction band. This ensures precise resonance alignment across the qubit array.

2. For each individual qubit, we perform quantum state tomography after Hamiltonian evolution under the required coupling strength of the couplers, while maintaining all other qubits detuned by approximately $\pm$ 100 MHz from their resonant frequencies. This configuration enables accurate experimental tracking of phase accumulation induced by reference frame transitions. We then finely adjust the Z-pulse amplitude (zpa) for each qubit until the accumulated phase satisfies the relation:
\begin{equation}\label{suppeq:addtionalphase}
\Delta \phi = (\omega_\text{resonance}-\omega_\text{gates})\times t\,.
\end{equation}
3. The additional phase for the $i$th qubit is calculated as $(\omega^i_\text{resonance}-\omega^i_\text{gates})\times t_\text{circuit}$, where $t_\text{circuit}= 228 $ ns. This phase correction is applied via single-qubit Z-rotations both before and after the analog evolution segment. Since our gate sequence consists of a layer of single-qubit gates followed by a layer of two-qubit gates, this phase correction can be efficiently implemented via virtual Z gates—that is, by adjusting the phase of subsequent XY pulses—rather than applying physical Z rotations.

4. Finally, we characterize and compensate for residual relative global phases between all qubit pairs in a sequential chain from the first to the last qubit, by introducing corresponding initial phase offsets in subsequent XY rotation pulses.

\section{Alternative optimization scheme}

In the main text, the imbalance observable $\mathcal{I}_{\bm z}$ was measured by applying the inverse circuit $\hat U_\mathcal{M}^\dagger(\bm z)$ before readout:
\begin{equation*}
\begin{quantikz}[column sep=0.2cm, row sep=0.25cm]
   \lstick{A} & \gate[2]{-\phi_1} & \qw                  &   \\
   \lstick{B}  &                    & \gate[2]{-\theta_1} & \meter{$S_B^z$}  \\
   \lstick{C}  & \gate[2]{-\phi_2} &                      & \meter{$S_C^z$}  \\
   \lstick{D}  &                    & \qw                 & 
\end{quantikz},\quad
\begin{quantikz}[column sep=0.2cm, row sep=0.25cm]
   \lstick{C}  & \gate[2]{-\phi_2} & \qw                  &   \\
   \lstick{D}  &                    & \gate[2]{-\theta_2} & \meter{$S_D^z$}  \\
   \lstick{A}  & \gate[2]{-\phi_1} &                      & \meter{$S_A^z$}  \\
   \lstick{B}  &                    & \qw                 & 
\end{quantikz}
\end{equation*}
This protocol directly probes the overlap between the evolved state and the variational ansatz parametrized by $\bm z$, and thus provides a practical proxy for the state fidelity. However, in practice, scanning the imbalance over many parameter choices $\bm z$ requires separate state preparation and reverse-gate operation for each setting, which quickly becomes resource-intensive. To mitigate this overhead, it is desirable to have an alternative scheme in which all measurements are performed in a fixed basis, while the results for arbitrary $\bm z$ can be reconstructed from a common set of measurements with analytically known coefficients. Such a scheme can be obtained by carrying out the evolution of an observable in the Heisenberg picture rather than evolving the state. 

Specifically, we consider
\begin{equation}
\begin{aligned}
    \hat O_1(\theta_1,\phi_1,\phi_2) &= U^\dagger(\theta_1,\phi_1,\phi_2) (S^z_B-S^z_C) U(\theta_1,\phi_1,\phi_2), \\
    \hat O_2(\theta_2,\phi_1,\phi_2) &= U^\dagger(\theta_2,\phi_1,\phi_2) (S^z_D-S^z_A) U(\theta_2,\phi_1,\phi_2),
\end{aligned}
\end{equation}
whose dependence on $\bm z$ can be determined analytically within the variational manifold. Once these operators are expressed as a linear combination of a fixed set of local correlators, the imbalance for any $\bm z$ follows from a weighted sum of those correlators, all measurable within a single experimental configuration. In the following, we derive the explicit form of the rotated operators $\tilde{\sigma}^z_j(\bm z)$ and show how the resulting linear combinations reproduce the imbalance landscape without requiring separate state re-preparation for each parameter value.

The explicit form of the observables are:
\begin{equation}
\begin{aligned}
\hat O_1(\theta_1,\phi_1,\phi_2) =&\ \cos(2\theta_1) \left(\sin^2\phi_1 Z_A+\cos ^2\phi_1 Z_B - \cos ^2\phi_2 Z_C - \sin ^2\phi_2 Z_D \right) \\
&+\cos(2 \theta _1) \left[
 \sin(2\phi_1) X_AY_B + \sin(2\phi_2) X_CY_D
\right] \\
&-2\sin(2\theta_1)\left(\cos\phi _1 \cos\phi _2 X_BY_C + \sin\phi_1 \sin\phi _2 X_A Z_B Z_C Y_D\right) \\
&-2\sin(2\theta_1) \left(
	\sin\phi_1 \cos\phi_2 X_A Z_B X_C 
	-\sin\phi_2 \cos\phi_1 X_B Z_C X_D 
	\right). \\
\hat O_2(\theta_2,\phi_1,\phi_2) =&\ \cos(2\theta_2) \left(\sin^2\phi_2 Z_C+\cos ^2\phi_2 Z_D - \cos ^2\phi_1 Z_A - \sin ^2\phi_1 Z_B \right) \\
&+\cos(2 \theta _2) \left[
 \sin(2\phi_2) X_CY_D + \sin(2\phi_1) X_AY_B
\right] \\
&-2\sin(2\theta_2)\left(\cos\phi _1 \cos\phi _2 X_DY_A + \sin\phi_1 \sin\phi _2 X_C Z_D Z_A Y_B\right) \\
&-2\sin(2\theta_2) \left(
	\sin\phi_2 \cos\phi_1 X_C Z_D X_A 
	-\sin\phi_1 \cos\phi_2 X_D Z_A X_B 
	\right).
\end{aligned}
\end{equation}
Note that we have used the symmetric definition for operators $X$ and $Y$: an $X_i X_j$ pair always means $(X_i X_j + Y_i Y_j)/2$, and an $X_i Y_j$ pair always means $(X_i Y_j - Y_i X_j)/2$.
When $\theta_1=\theta_2=\theta$, the sum of two operators is:
\begin{equation}
\begin{aligned}
[\hat O_1+\hat O_2](\theta,\phi_1,\phi_2) =&\ -\cos(2\theta)\left[\cos(2\phi_1)(Z_A-Z_B)+\cos(2\phi_2)(Z_C-Z_D)\right] \\
&+2\cos(2\theta) \left[\sin(2\phi_1) X_AY_B +\sin(2\phi_2) X_CY_D\right] \\
&-2\sin(2\theta)\cos\phi _1 \cos\phi _2 (X_BY_C+X_DY_A) \\
&-2\sin(2\theta)\sin\phi_1 \sin\phi_2 (X_A Z_B Z_C Y_D+X_C Z_D Z_A Y_B) \\
&+2\sin(2\theta)\sin\phi_1\cos\phi_2 (X_D Z_A X_B - X_A Z_B X_C) \\
&+2\sin(2\theta)\sin\phi_2\cos\phi_1 (X_B Z_C X_D - X_C Z_D X_A).
\end{aligned}
\end{equation}
These two local observables are combinations of fixed local operators, with specific coefficients depending on $\theta,\phi$.

Experimentally, given a quantum state, the variational parameters of the ansatz can be inferred by maximizing the expectation value of suitably chosen local observables. In particular,
\begin{equation}
    \{\theta, \phi_1, \phi_2\}
    = \underset{\theta, \phi_1, \phi_2}{\operatorname{argmax}}
    \left[ \langle \hat O_1 + \hat O_2 \rangle(\theta, \phi_1, \phi_2) \right],
\end{equation}
where $\hat O_{1,2}$ denote experimentally accessible operators such as local imbalances or spin correlators. 

This procedure plays a key role in the hybrid feedback protocol: it transforms the quantum-classical optimization step into a purely classical optimization problem. Rather than reconstructing the full quantum state multiple times, one only measures the expectation values of a few local observables and uses their classical post-processing to update the parameters $(\theta, \phi_1, \phi_2)$. Consequently, the feedback loop alternates between short quantum evolutions and entirely classical optimization steps, minimizing experimental overhead while retaining full access to the relevant variational degrees of freedom.

\section{Time-dependent variational principle for shallow circuit ansatz}
\label{apx:mps_ansatz}

We begin by constructing a variational ansatz based on nearest-neighbor XY rotations applied to the N\'eel-ordered product state $\ket{\mathbb{Z}_2}\equiv |01\dots01\rangle$, which serves as a simple unentangled reference with alternating spin order. The ansatz captures the essential entangling dynamics of the interacting SSH ladder through layers of XY rotation gates, each of which acts on a pair of qubits and preserves total magnetization. The two-qubit rotation is defined as
\begin{equation}
	\operatorname{XY}(\theta)=
	\begin{bmatrix}
		1 & 0 & 0 & 0 \\
		0 & \cos\theta & i\sin \theta & 0 \\
		0 & i\sin \theta & \cos\theta & 0 \\
		0 & 0 & 0 & 1
	\end{bmatrix},
\end{equation}
which generates coherent oscillations in the single-excitation subspace $\{|01\rangle,|10\rangle\}$.
Applying layers of such gates with alternating parameters produces the variational state
\begin{equation}
\label{apxeq:mps_ansatz}
	|\psi(\theta_1,\theta_2,\phi_1,\phi_2)\rangle 
	= \left[\prod_{j} \operatorname{XY}_{4j+1,4j+2}(\phi_1)\,
	\operatorname{XY}_{4j+3,4j+4}(\phi_2)\right]
	\times \left[\prod_{j} \operatorname{XY}_{4j,4j+1}(\theta_1)\,
	\operatorname{XY}_{4j+2,4j+3}(\theta_2)\right] |\mathbb Z_2\rangle.
\end{equation}
This ansatz admits a natural matrix-product state (MPS) representation. A schematic tensor-network depiction is:
\begin{equation}
\label{apxeq:mps_ansatz}
	|\psi\rangle 
	= \begin{tikzpicture}[tensornetwork]
        \node[atensor] (A1) at (0, 0) {$A_{\phi_1}$};
        \node[atensor] (A2) at (1.25, 0) {$B_{\theta_1}$};
        \node[atensor] (A3) at (2.5, 0) {$C_{\phi_2}$};
        \node[atensor] (A4) at (3.75, 0) {$D_{\theta_2}$};
        \draw (A1.north) -- +(0, +0.5);
        \draw (A2.north) -- +(0, +0.5);
        \draw (A3.north) -- +(0, +0.5);
        \draw (A4.north) -- +(0, +0.5);
        \draw (A1) -- (A2) -- (A3) -- (A4);
        \draw (A1.west) -- +(-0.3, 0);
        \draw (A4.east) -- +(0.3, 0);
    \end{tikzpicture}.
\end{equation}
Each tensor is parametrized explicitly by the gate angles:
\begin{equation}
\begin{aligned}
A_{\phi_1}^{[0]} &= \begin{bmatrix}
 0 & 0 & i \sin (\phi _1) & 0 \\
 \cos ^2\left(\frac{\phi _1}{2}\right) & \sin ^2\left(\frac{\phi _1}{2}\right) & 0 & 0
\end{bmatrix}, &
A_{\phi_1}^{[1]} &= \begin{bmatrix}
 \cos ^2\left(\frac{\phi _1}{2}\right) & -\sin ^2\left(\frac{\phi _1}{2}\right) & 0 & 0 \\
 0 & 0 & 0 & i \sin (\phi _1) 
\end{bmatrix}, \\
B_{\theta_1}^{[0]} &= \begin{bmatrix}
	\cos (\theta _1) & 0 \\
	\cos (\theta _1) & 0 \\
	0 & 0 \\
	0 & i \sin (\theta _1)
\end{bmatrix}, & B_{\theta_1}^{[1]} &= \begin{bmatrix}
	0 & i \sin (\theta _1) \\
 0 & -i \sin (\theta _1) \\
 \cos (\theta _1) & 0 \\
 0 & 0
\end{bmatrix}, \\
C^{[i]}_{\phi_2} &= A^{[i]}_{\phi_1 \rightarrow \phi_2}, &
D^{[i]}_{\theta_2} &= B^{[i]}_{\theta_1 \rightarrow \theta_2}.
\end{aligned}
\end{equation}
Expectation values and overlaps can be evaluated using the transfer matrices formed by contracting each tensor with its conjugate:
\begin{equation}
\begin{aligned}
    \begin{tikzpicture}[tensornetwork]
        \node[atensor] (A1) at (0, 1) {$\bar A_{\phi_1}$};
        \node[atensor] (A2) at (1.5, 1) {$\bar B_{\theta_1}$};
        \node[atensor] (A3) at (0, -1) {$A_{\phi_1}$};
        \node[atensor] (A4) at (1.5, -1) {$B_{\theta_1}$};
        \draw (A1) -- (A2);
        \draw (A1) -- (A3);
        \draw (A2) -- (A4);
        \draw (A3) -- (A4);
        \draw (A1.west) -- +(-0.3,0);
        \draw (A3.west) -- +(-0.3,0);
        \draw (A2.east) -- +(0.3,0);
        \draw (A4.east) -- +(0.3,0);
    \end{tikzpicture}
    &= \begin{bmatrix}
		\cos ^2 \theta_1 & 0 & 0 & \sin ^2\theta_1 \\
 		0 & 0 & 0 & 0 \\
 		0 & 0 & 0 & 0 \\
 		\cos ^2\theta_1 & 0 & 0 & \sin ^2\theta_1
	\end{bmatrix}
	= \begin{tikzpicture}[tensornetwork]
        \draw[rounded corners] (0, 1.25) -- +(0.5, 0) -- +(0.5, -2.5) -- +(0, -2.5);
    \end{tikzpicture}
    \begin{tikzpicture}[tensornetwork]
        \coordinate (A1) at (0, 1.25) {};
        \coordinate (A1conj) at (0, -1.25) {};
        \node[ctensor, label={left, text depth=}:] (rho) at (0, 0) {$L_{\theta_1}$};
        \draw[rounded corners] (rho.north) -- (rho.north |- A1.west) -- +(0.5, 0);
        \draw[rounded corners] (rho.south) -- (rho.south |- A1conj.west) -- +(0.5, 0);
    \end{tikzpicture}, \\
    \begin{tikzpicture}[tensornetwork]
        \node[atensor] (A1) at (0, 1) {$\bar C_{\phi_2}$};
        \node[atensor] (A2) at (1.5, 1) {$\bar D_{\theta_2}$};
        \node[atensor] (A3) at (0, -1) {$C_{\phi_2}$};
        \node[atensor] (A4) at (1.5, -1) {$D_{\theta_2}$};
        \draw (A1) -- (A2);
        \draw (A1) -- (A3);
        \draw (A2) -- (A4);
        \draw (A3) -- (A4);
        \draw (A1.west) -- +(-0.3,0);
        \draw (A3.west) -- +(-0.3,0);
        \draw (A2.east) -- +(0.3,0);
        \draw (A4.east) -- +(0.3,0);
    \end{tikzpicture}
    &= \begin{bmatrix}
		\cos ^2 \theta_2 & 0 & 0 & \sin ^2\theta_2 \\
 		0 & 0 & 0 & 0 \\
 		0 & 0 & 0 & 0 \\
 		\cos ^2\theta_2 & 0 & 0 & \sin ^2\theta_2
	\end{bmatrix}
	= \begin{tikzpicture}[tensornetwork]
        \draw[rounded corners] (0, 1.25) -- +(0.5, 0) -- +(0.5, -2.5) -- +(0, -2.5);
    \end{tikzpicture}
    \begin{tikzpicture}[tensornetwork]
        \coordinate (A1) at (0, 1.25) {};
        \coordinate (A1conj) at (0, -1.25) {};
        \node[ctensor, label={left, text depth=}:] (rho) at (0, 0) {$L_{\theta_2}$};
        \draw[rounded corners] (rho.north) -- (rho.north |- A1.west) -- +(0.5, 0);
        \draw[rounded corners] (rho.south) -- (rho.south |- A1conj.west) -- +(0.5, 0);
    \end{tikzpicture}.
\end{aligned}
\end{equation}
The corresponding left dominant eigenvector is:
\begin{equation}
	L_\theta = \begin{bmatrix}
		\cos^2\theta & 0 \\
		0 & \sin^2\theta
	\end{bmatrix}.
\end{equation}

Time-dependent variational principle (TDVP) provides a systematic way to approximate the real-time evolution 
of a quantum state by restricting the dynamics to a chosen variational manifold. Concretely, the 
Schr\"odinger equation
\begin{equation}
    i \frac{\partial}{\partial t} |\psi({\bm z})\rangle = \hat H |\psi({\bm z})\rangle
\end{equation}
is projected onto the tangent space of the manifold spanned by the variational parameters ${\bm z}=\{\theta_1,\theta_2,\phi_1,\phi_2\}$. 
This projection ensures that the time evolution of the parameters follows the trajectory within the manifold that minimizes the instantaneous distance between the exact and variational time derivatives of the state. 
As a result, the linear quantum dynamics are transformed into a set of nonlinear coupled equations of motion for the parameters, which can display rich dynamical features including periodic motion, instability, and chaos.

In general, the TDVP equations of motion take the form 
\begin{equation}
\sum_{l} G_{kl}({\bm z}) \, \dot z_l = - i \, h_k({\bm z}),
\label{eq:general_tdvp}
\end{equation}
where the Gram matrix is defined as
\begin{equation}
    G_{kl}({\bm z}) = \langle \partial_{z_k}\psi({\bm z}) \,|\, \partial_{z_l}\psi({\bm z})\rangle,
\end{equation}
and the Hamiltonian gradient is given by
\begin{equation}
    h_k({\bm z}) = \langle \partial_{z_k}\psi({\bm z}) \,|\, \hat H \,|\, \psi({\bm z})\rangle.
\end{equation}
The expression~\eqref{eq:general_tdvp} represents a system of linear equations for the parameter velocities $\dot z_l$, 
with coefficients determined entirely by the geometry of the variational manifold and the action of the Hamiltonian. In the following, we derive the explicit form of these equations for the XY-rotation ansatz introduced above. 
We begin by computing the Gram matrix and Hamiltonian gradient, before arriving at closed-form expressions for the equations of motion of the variational parameters $(\theta_1,\theta_2,\phi_1,\phi_2)$.

\subsection{Gram matrix}

The Gram matrix encodes the overlaps of tangent vectors $\{|\partial_{z_\mu}\psi\rangle\}$ on the variational manifold,
\[
G_{\mu\nu} \equiv \langle \partial_{z_\mu}\psi | \partial_{z_\nu}\psi\rangle,
\qquad
z_\mu \in \{\theta_1,\theta_2,\phi_1,\phi_2\}.
\]
Throughout this section, the prime symbol on a tensor denotes the derivative with respect to the corresponding parameter (e.g., $A' \equiv \partial_{\phi_1}A$, $B' \equiv \partial_{\theta_1}B$). 
The symbols $L_1$ and $L_2$ denote the left fixed points of the transfer matrices associated with the
$(C,D)$ and $(A,B)$ layers, respectively.

By contracting the relevant two-layer tensor networks with one insertion of a tangent tensor on the bra/ket legs,
the diagonal entries are
\begin{eqnarray}
G_{\phi_1,\phi_1} &=&
	\begin{tikzpicture}[tensornetwork]
        \node[atensor] (A1) at (0, 1) {$\bar A'$};
        \node[atensor] (A2) at (1, 1) {$\bar{B}$};
        \node[atensor] (A3) at (0, -1) {$A'$};
        \node[atensor] (A4) at (1, -1) {$B$};
        \node[atensor] (L2) at (-0.7,0) {$L_2$};
        \draw (A1) -- (A2);
        \draw (A1) -- (A3);
        \draw (A2) -- (A4);
        \draw (A3) -- (A4);
        \draw[rounded corners] (L2.north) -- (-0.7,1) -- (A1.west);
        \draw[rounded corners] (L2.south) -- (-0.7,-1) -- (A3.west);
        \draw[rounded corners] (A2.east) -- +(0.3,0) -- +(0.3,-2) -- (A4.east);
    \end{tikzpicture}
    = \frac{1+\cos(2\theta_1)\cos(2\theta_2)}{2}, \quad
	G_{\theta_1,\theta_1} =
	\begin{tikzpicture}[tensornetwork]
        \node[atensor] (A1) at (0, 1) {$\bar A$};
        \node[atensor] (A2) at (1, 1) {$\bar{B}'$};
        \node[atensor] (A3) at (0, -1) {$A$};
        \node[atensor] (A4) at (1, -1) {$B'$};
        \node[atensor] (L2) at (-0.7,0) {$L_2$};
        \draw (A1) -- (A2);
        \draw (A1) -- (A3);
        \draw (A2) -- (A4);
        \draw (A3) -- (A4);
        \draw[rounded corners] (L2.north) -- (-0.7,1) -- (A1.west);
        \draw[rounded corners] (L2.south) -- (-0.7,-1) -- (A3.west);
        \draw[rounded corners] (A2.east) -- +(0.3,0) -- +(0.3,-2) -- (A4.east);
    \end{tikzpicture}=1, \\
    G_{\phi_2,\phi_2} &=&
	\begin{tikzpicture}[tensornetwork]
        \node[atensor] (A1) at (0, 1) {$\bar C'$};
        \node[atensor] (A2) at (1, 1) {$\bar{D}$};
        \node[atensor] (A3) at (0, -1) {$C'$};
        \node[atensor] (A4) at (1, -1) {$D$};
        \node[atensor] (L2) at (-0.7,0) {$L_1$};
        \draw (A1) -- (A2);
        \draw (A1) -- (A3);
        \draw (A2) -- (A4);
        \draw (A3) -- (A4);
        \draw[rounded corners] (L2.north) -- (-0.7,1) -- (A1.west);
        \draw[rounded corners] (L2.south) -- (-0.7,-1) -- (A3.west);
        \draw[rounded corners] (A2.east) -- +(0.3,0) -- +(0.3,-2) -- (A4.east);
    \end{tikzpicture}
    = \frac{1+\cos(2\theta_1)\cos(2\theta_2)}{2},\quad
	G_{\theta_2,\theta_2} =
	\begin{tikzpicture}[tensornetwork]
        \node[atensor] (A1) at (0, 1) {$\bar C$};
        \node[atensor] (A2) at (1, 1) {$\bar{D}'$};
        \node[atensor] (A3) at (0, -1) {$C$};
        \node[atensor] (A4) at (1, -1) {$D'$};
        \node[atensor] (L2) at (-0.7,0) {$L_1$};
        \draw (A1) -- (A2);
        \draw (A1) -- (A3);
        \draw (A2) -- (A4);
        \draw (A3) -- (A4);
        \draw[rounded corners] (L2.north) -- (-0.7,1) -- (A1.west);
        \draw[rounded corners] (L2.south) -- (-0.7,-1) -- (A3.west);
        \draw[rounded corners] (A2.east) -- +(0.3,0) -- +(0.3,-2) -- (A4.east);
    \end{tikzpicture}=1.
\end{eqnarray}
All off-diagonal elements vanish, $G_{\mu\nu}=0$ for $\mu\neq \nu$.

\subsection{Hamiltonian gradients}

In this subsection, we calculate the explicit form of the Hamiltonian gradients part of the TDVP equation of motion:
\begin{equation}
    h_\mu \equiv \langle \partial_{z_\mu}\psi | \hat H | \psi\rangle,
\qquad z_\mu\in\{\theta_1,\theta_2,\phi_1,\phi_2\}.
\end{equation}
The Hamiltonian can be split into an SSH part $\hat H_0$ and a longer–range correction $\hat V$,
\begin{equation}
\begin{aligned}
    \hat H&=\hat H_0+\hat V,\\ 
    \hat H_0&=-\sum_j\big(J_o\,\hat\sigma^+_{2j-1}\hat\sigma^-_{2j}
+J_e\,\hat\sigma^+_{2j}\hat\sigma^-_{2j+1}+\mathrm{h.c.}\big),\\
    \hat V&=-J_3\sum_j\big(\hat\sigma^+_{2j-1}\hat\sigma^-_{2j+2}+\mathrm{h.c.}\big).
\end{aligned}
\end{equation}
For clarity, we separate the contributions of $\hat H_0$ and $\hat V$ and use tensor-network diagrams to indicate the corresponding local operator insertions. We denote by $h_o$, $h_e$, and $h_3$ the local MPO tensors representing the odd-rung, even-bond, and next-nearest interactions, respectively; $L_1$ and $L_2$ are the left fixed points of the transfer matrices for the $(C,D)$ and $(A,B)$ layers.

\paragraph{SSH contribution.}
The three nonvanishing gradients from $\hat H_0$ are obtained by inserting
$h_o$ or $h_e$ in the appropriate two-layer contractions. In diagrammatic form:
\begin{eqnarray*}
	-i\langle \partial_{\phi_1}\psi|\hat H_0|\psi\rangle &=&
	\begin{tikzpicture}[tensornetwork]
        \node[atensor]  (Dd) at (0, 1) {$\bar A'$};
        \node[atensor]  (Ad) at (1, 1) {$\bar B$};
        \node[atensor]  (D) at (0, -1) {$A$};
        \node[atensor]  (A) at (1, -1) {$B$};
        \node[widetensor=2] (O) at (0.5, 0) {$h_o$};
        \node[atensor] (L2) at (-0.7,0) {$L_2$};
        \draw (Dd) --(Ad);
        \draw (A) -- (D);
        \draw (Dd) -- (O.north -| Dd);
        \draw (D) -- (O.south -| D);
        \draw (Ad) -- (O.north -| Ad);
        \draw (A) -- (O.south -| A);
        \draw[rounded corners] (L2.north) -- (-0.7, 1) -- (Dd.west);
        \draw[rounded corners] (L2.south) -- (-0.7, -1) -- (D.west);
        \draw[rounded corners] (Ad.east) -- +(0.3, 0) -- +(0.3, -2) -- (A.east);
    \end{tikzpicture}
    +\begin{tikzpicture}[tensornetwork]
        \node[atensor] (Dd) at (0, 1) {$\bar C$};
        \node[atensor] (Ad) at (1, 1) {$\bar D$};
        \node[atensor] (Bd) at (2, 1) {$\bar A'$};
        \node[atensor] (Cd) at (3, 1) {$\bar B$};
        \node[atensor] (D) at (0, -1) {$C$};
        \node[atensor] (A) at (1, -1) {$D$};
        \node[atensor] (B) at (2, -1) {$A$};
        \node[atensor] (C) at (3, -1) {$B$};
        \node[widetensor=2] (O) at (1.5, 0) {$h_e$};
        \node[atensor] (L2) at (-0.7,0) {$L_1$};
        \draw (Dd) -- (D);
        \draw (Ad) -- (O.north -| Ad);
        \draw (A) -- (O.south -| A);
        \draw (Bd) -- (O.north -| Bd);
        \draw (B) -- (O.south -| B);
        \draw (Cd) -- (C);
        \draw (Dd) -- (Ad);
        \draw (D) -- (A);
        \draw (Ad) -- (Bd);
        \draw (A) -- (B);
        \draw (Bd) -- (Cd);
        \draw (B) -- (C);
        \draw[rounded corners] (L2.north) -- (-0.7,1) -- (Dd.west);
        \draw[rounded corners] (L2.south) -- (-0.7,-1) -- (D.west);
        \draw[rounded corners] (Cd.east) -- +(0.3,0) -- +(0.3,-2) -- (C.east);
    \end{tikzpicture}
    +\begin{tikzpicture}[tensornetwork]
        \node[atensor] (Bd) at (0, 1) {$\bar A'$};
        \node[atensor] (Cd) at (1, 1) {$\bar B$};
        \node[atensor] (Dd) at (2, 1) {$\bar C$};
        \node[atensor] (Ad) at (3, 1) {$\bar D$};
        \node[atensor] (B) at (0, -1) {$A$};
        \node[atensor] (C) at (1, -1) {$B$};
        \node[atensor] (D) at (2, -1) {$C$};
        \node[atensor] (A) at (3, -1) {$D$};
        \node[widetensor=2] (O) at (1.5, 0) {$h_e$};
        \node[atensor] (L1) at (-0.7,0) {$L_2$};
        \draw (Bd) -- (B);
        \draw (Cd) -- (O.north -| Cd);
        \draw (C) -- (O.south -| C);
        \draw (Dd) -- (O.north -| Dd);
        \draw (D) -- (O.south -| D);
        \draw (Ad) -- (A);
        \draw (Dd) -- (Ad);
        \draw (D) -- (A);
        \draw (Cd) -- (Bd);
        \draw (C) -- (B);
        \draw (Dd) -- (Cd);
        \draw (D) -- (C);
        \draw[rounded corners] (L1.north) -- (-0.7,1) -- (Bd.west);
        \draw[rounded corners] (L1.south) -- (-0.7,-1) -- (B.west);
        \draw[rounded corners] (Ad.east) -- +(0.3,0) -- +(0.3,-2) -- (A.east);
    \end{tikzpicture} \nonumber \\
	&=& \frac{J_o}{2}\left[1+\cos \left(2 \theta _1\right) \cos \left(2 \theta _2\right)\right]+ \frac{J_e}{2}\sin(2 \theta _1+2\theta _2) \sin \left(\phi _1\right) \cos \left(\phi _2\right), \\
	-i\langle \partial_{\phi_2}\psi|\hat H_0|\psi\rangle 
	&=& \left. i\langle \partial_{\phi_1}\psi|\hat H_0|\psi\rangle \right|^{\theta_1\leftrightarrow\theta_2}_{\phi_1\leftrightarrow\phi_2} 
	= \frac{J_o}{2}\left[1+\cos \left(2 \theta _1\right) \cos \left(2 \theta _2\right)\right]+ \frac{J_e}{2}\sin(2 \theta _1+2\theta _2) \sin \left(\phi _2\right) \cos \left(\phi _1\right), \\
	-i\langle \partial_{\theta_1}\psi|\hat H_0|\psi\rangle 
	&=&\begin{tikzpicture}[tensornetwork]
        \node[atensor]  (Dd) at (0, 1) {$\bar A$};
        \node[atensor]  (Ad) at (1, 1) {$\bar B'$};
        \node[atensor]  (D) at (0, -1) {$A$};
        \node[atensor]  (A) at (1, -1) {$B$};
        \node[widetensor=2] (O) at (0.5, 0) {$h_o$};
        \node[atensor] (L2) at (-0.7,0) {$L_2$};
        \draw (Dd) --(Ad);
        \draw (A) -- (D);
        \draw (Dd) -- (O.north -| Dd);
        \draw (D) -- (O.south -| D);
        \draw (Ad) -- (O.north -| Ad);
        \draw (A) -- (O.south -| A);
        \draw[rounded corners] (L2.north) -- (-0.7, 1) -- (Dd.west);
        \draw[rounded corners] (L2.south) -- (-0.7, -1) -- (D.west);
        \draw[rounded corners] (Ad.east) -- +(0.3, 0) -- +(0.3, -2) -- (A.east);
    \end{tikzpicture}
    +\begin{tikzpicture}[tensornetwork]
        \node[atensor] (Dd) at (0, 1) {$\bar A$};
        \node[atensor] (Ad) at (1, 1) {$\bar B'$};
        \node[atensor] (Bd) at (2, 1) {$\bar C$};
        \node[atensor] (Cd) at (3, 1) {$\bar D$};
        \node[atensor] (D) at (0, -1) {$A$};
        \node[atensor] (A) at (1, -1) {$B$};
        \node[atensor] (B) at (2, -1) {$C$};
        \node[atensor] (C) at (3, -1) {$D$};
        \node[widetensor=2] (O) at (1.5, 0) {$h_e$};
        \node[atensor] (L2) at (-0.7,0) {$L_2$};
        \draw (Dd) -- (D);
        \draw (Ad) -- (O.north -| Ad);
        \draw (A) -- (O.south -| A);
        \draw (Bd) -- (O.north -| Bd);
        \draw (B) -- (O.south -| B);
        \draw (Cd) -- (C);
        \draw (Dd) -- (Ad);
        \draw (D) -- (A);
        \draw (Ad) -- (Bd);
        \draw (A) -- (B);
        \draw (Bd) -- (Cd);
        \draw (B) -- (C);
        \draw[rounded corners] (L2.north) -- (-0.7,1) -- (Dd.west);
        \draw[rounded corners] (L2.south) -- (-0.7,-1) -- (D.west);
        \draw[rounded corners] (Cd.east) -- +(0.3,0) -- +(0.3,-2) -- (C.east);
    \end{tikzpicture}
    +\begin{tikzpicture}[tensornetwork]
        \node[atensor] (Bd) at (0, 1) {$\bar C$};
        \node[atensor] (Cd) at (1, 1) {$\bar D$};
        \node[atensor] (Dd) at (2, 1) {$\bar A$};
        \node[atensor] (Ad) at (3, 1) {$\bar B'$};
        \node[atensor] (B) at (0, -1) {$C$};
        \node[atensor] (C) at (1, -1) {$D$};
        \node[atensor] (D) at (2, -1) {$A$};
        \node[atensor] (A) at (3, -1) {$B$};
        \node[widetensor=2] (O) at (1.5, 0) {$h_e$};
        \node[atensor] (L1) at (-0.7,0) {$L_1$};
        \draw (Bd) -- (B);
        \draw (Cd) -- (O.north -| Cd);
        \draw (C) -- (O.south -| C);
        \draw (Dd) -- (O.north -| Dd);
        \draw (D) -- (O.south -| D);
        \draw (Ad) -- (A);
        \draw (Dd) -- (Ad);
        \draw (D) -- (A);
        \draw (Cd) -- (Bd);
        \draw (C) -- (B);
        \draw (Dd) -- (Cd);
        \draw (D) -- (C);
        \draw[rounded corners] (L1.north) -- (-0.7,1) -- (Bd.west);
        \draw[rounded corners] (L1.south) -- (-0.7,-1) -- (B.west);
        \draw[rounded corners] (Ad.east) -- +(0.3,0) -- +(0.3,-2) -- (A.east);
    \end{tikzpicture} 
	= J_e \cos (\phi _1) \cos (\phi _2), \\
	-i\langle \partial_{\theta_2}\psi|\hat H_0|\psi\rangle 
	&=& -\left. i\langle \partial_{\theta_1}\psi|\hat H_0|\psi\rangle \right|^{\theta_1\leftrightarrow\theta_2}_{\phi_1\leftrightarrow\phi_2}
	= J_e \cos (\phi _1) \cos (\phi _2).
\end{eqnarray*}

\paragraph{Interaction contribution.}
The next-nearest interaction is encoded by two insertions of $h_3$ bridged across a layer. The corresponding gradients are
\begin{eqnarray*}
	-i\langle \partial_{\phi_1}\psi|\hat V|\psi\rangle &=&  
	\begin{tikzpicture}[tensornetwork]
        \node[atensor] (Bd) at (0, 1) {$\bar A'$};
        \node[atensor] (Cd) at (1, 1) {$\bar B$};
        \node[atensor] (Dd) at (2, 1) {$\bar C$};
        \node[atensor] (Ad) at (3, 1) {$\bar D$};
        \node[atensor] (B) at (0, -1) {$A$};
        \node[atensor] (C) at (1, -1) {$B$};
        \node[atensor] (D) at (2, -1) {$C$};
        \node[atensor] (A) at (3, -1) {$D$};
        \node[atensor] (L1) at (-0.7,0) {$L_2$};
        \node[tensor] (O) at (0, 0) {$h_3$};
        \node[tensor] (O2) at (3, 0) {$h_3$};
        \draw (Cd) -- (C);
        \draw (Bd) -- (O.north -| Bd);
        \draw (B) -- (O.south -| B);
        \draw (Ad) -- (O2.north -| Ad);
        \draw (A) -- (O2.south -| A);
        \draw (Dd) -- (D);
        \draw (Dd) -- (Ad);
        \draw (D) -- (A);
        \draw (Cd) -- (Bd);
        \draw (C) -- (B);
        \draw (Dd) -- (Cd);
        \draw (D) -- (C);
        \draw (O.east) -- (O2.west);
        \draw[rounded corners] (L1.north) -- (-0.7,1) -- (Bd.west);
        \draw[rounded corners] (L1.south) -- (-0.7,-1) -- (B.west);
        \draw[rounded corners] (Ad.east) -- +(0.3,0) -- +(0.3,-2) -- (A.east);
    \end{tikzpicture} + 
	\begin{tikzpicture}[tensornetwork]
        \node[atensor] (Dd) at (0, 1) {$\bar C$};
        \node[atensor] (Ad) at (1, 1) {$\bar D$};
        \node[atensor] (Bd) at (2, 1) {$\bar A'$};
        \node[atensor] (Cd) at (3, 1) {$\bar B$};
        \node[atensor] (D) at (0, -1) {$C$};
        \node[atensor] (A) at (1, -1) {$D$};
        \node[atensor] (B) at (2, -1) {$A$};
        \node[atensor] (C) at (3, -1) {$B$};
        \node[atensor] (L2) at (-0.7,0) {$L_1$};
        \node[tensor] (O) at (0, 0) {$h_3$};
        \node[tensor] (O2) at (3, 0) {$h_3$};
        \draw (Ad) -- (A);
        \draw (Dd) -- (O.north -| Dd);
        \draw (D) -- (O.south -| D);
        \draw (Cd) -- (O.north -| Cd);
        \draw (C) -- (O.south -| C);
        \draw (Bd) -- (B);
        \draw (Dd) -- (Ad);
        \draw (D) -- (A);
        \draw (Ad) -- (Bd);
        \draw (A) -- (B);
        \draw (Bd) -- (Cd);
        \draw (B) -- (C);
        \draw (O.east) -- (O2.west);
        \draw[rounded corners] (L2.north) -- (-0.7,1) -- (Dd.west);
        \draw[rounded corners] (L2.south) -- (-0.7,-1) -- (D.west);
        \draw[rounded corners] (Cd.east) -- +(0.3,0) -- +(0.3,-2) -- (C.east);
    \end{tikzpicture} 
    = \frac{J_3}{2}  \sin \left(2 \theta _1+2\theta _2\right) \sin \left(\phi _2\right) \cos \left(\phi _1\right), \\
    -i\langle \partial_{\phi_2}\psi|\hat V|\psi\rangle 
    &=& -\left. i\langle \partial_{\phi_1}\psi|\hat V|\psi\rangle \right|^{\theta_1\leftrightarrow\theta_2}_{\phi_1\leftrightarrow\phi_2}
    =\frac{J_3}{2}  \sin \left(2 \theta _1+2\theta _2\right) \sin \left(\phi_1\right) \cos \left(\phi_2\right), \\
	-i\langle \partial_{\theta_1}\psi|\hat V|\psi\rangle &=&  
	\begin{tikzpicture}[tensornetwork]
        \node[atensor] (Dd) at (0, 1) {$\bar A$};
        \node[atensor] (Ad) at (1, 1) {$\bar B'$};
        \node[atensor] (Bd) at (2, 1) {$\bar C$};
        \node[atensor] (Cd) at (3, 1) {$\bar D$};
        \node[atensor] (D) at (0, -1) {$A$};
        \node[atensor] (A) at (1, -1) {$B$};
        \node[atensor] (B) at (2, -1) {$C$};
        \node[atensor] (C) at (3, -1) {$D$};
        \node[atensor] (L2) at (-0.7,0) {$L_2$};
        \node[tensor] (O) at (0, 0) {$h_3$};
        \node[tensor] (O2) at (3, 0) {$h_3$};
        \draw (Ad) -- (A);
        \draw (Dd) -- (O.north -| Dd);
        \draw (D) -- (O.south -| D);
        \draw (Cd) -- (O.north -| Cd);
        \draw (C) -- (O.south -| C);
        \draw (Bd) -- (B);
        \draw (Dd) -- (Ad);
        \draw (D) -- (A);
        \draw (Ad) -- (Bd);
        \draw (A) -- (B);
        \draw (Bd) -- (Cd);
        \draw (B) -- (C);
        \draw (O.east) -- (O2.west);
        \draw[rounded corners] (L2.north) -- (-0.7,1) -- (Dd.west);
        \draw[rounded corners] (L2.south) -- (-0.7,-1) -- (D.west);
        \draw[rounded corners] (Cd.east) -- +(0.3,0) -- +(0.3,-2) -- (C.east);
    \end{tikzpicture} +
    \begin{tikzpicture}[tensornetwork]
        \node[atensor] (Bd) at (0, 1) {$\bar C$};
        \node[atensor] (Cd) at (1, 1) {$\bar D$};
        \node[atensor] (Dd) at (2, 1) {$\bar A$};
        \node[atensor] (Ad) at (3, 1) {$\bar B'$};
        \node[atensor] (B) at (0, -1) {$C$};
        \node[atensor] (C) at (1, -1) {$D$};
        \node[atensor] (D) at (2, -1) {$A$};
        \node[atensor] (A) at (3, -1) {$B$};
        \node[atensor] (L1) at (-0.7,0) {$L_1$};
        \node[tensor] (O) at (0, 0) {$h_3$};
        \node[tensor] (O2) at (3, 0) {$h_3$};
        \draw (Cd) -- (C);
        \draw (Bd) -- (O.north -| Bd);
        \draw (B) -- (O.south -| B);
        \draw (Ad) -- (O2.north -| Ad);
        \draw (A) -- (O2.south -| A);
        \draw (Dd) -- (D);
        \draw (Dd) -- (Ad);
        \draw (D) -- (A);
        \draw (Cd) -- (Bd);
        \draw (C) -- (B);
        \draw (Dd) -- (Cd);
        \draw (D) -- (C);
        \draw (O.east) -- (O2.west);
        \draw[rounded corners] (L1.north) -- (-0.7,1) -- (Bd.west);
        \draw[rounded corners] (L1.south) -- (-0.7,-1) -- (B.west);
        \draw[rounded corners] (Ad.east) -- +(0.3,0) -- +(0.3,-2) -- (A.east);
    \end{tikzpicture} 
    = -J_3 \cos ^2\left(2 \theta _2\right) \sin \left(\phi _1\right) \sin \left(\phi _2\right), \\
    -i\langle \partial_{\theta_2}\psi|\hat V|\psi\rangle 
    &=& \left. i\langle \partial_{\theta_1}\psi|\hat V|\psi\rangle \right|^{\theta_1\leftrightarrow\theta_2}_{\phi_1\leftrightarrow\phi_2}
    =-J_3 \cos ^2\left(2 \theta _1\right) \sin \left(\phi _1\right) \sin \left(\phi _2\right).
\end{eqnarray*}
Collecting the SSH and interaction pieces, the Hamiltonian gradients entering the TDVP read
\begin{eqnarray}
-\,i\langle \partial_{\theta_1}\psi\mid \hat H\mid \psi\rangle
&=& J_e\,\cos\phi_1\cos\phi_2 - J_3\,\cos^2(2\theta_2)\,\sin\phi_1\sin\phi_2,\\
-\,i\langle \partial_{\theta_2}\psi\mid \hat H\mid \psi\rangle
&=& J_e\,\cos\phi_1\cos\phi_2 - J_3\,\cos^2(2\theta_1)\,\sin\phi_1\sin\phi_2, \\
-\,i\langle \partial_{\phi_1}\psi\mid \hat H\mid \psi\rangle
&=&\frac{J_o}{2}\!\left[1+\cos(2\theta_1)\cos(2\theta_2)\right]
+\frac{\sin(2\theta_1{+}2\theta_2)}{2}\,\big(J_e\,\sin\phi_1\cos\phi_2+J_3\,\sin\phi_2\cos\phi_1\big),\\
-\,i\langle \partial_{\phi_2}\psi\mid \hat H\mid \psi\rangle
&=&\frac{J_o}{2}\!\left[1+\cos(2\theta_1)\cos(2\theta_2)\right]
+\frac{\sin(2\theta_1{+}2\theta_2)}{2}\,\big(J_e\,\sin\phi_2\cos\phi_1+J_3\,\sin\phi_1\cos\phi_2\big).
\end{eqnarray}

\subsection{Equations of motion}

With the explicit form of the Gram matrix and Hamiltonian gradients, we obtain the TDVP equations of motion:
 \begin{equation}
 \begin{aligned}
         \dot \theta_1 &= J_e \cos \left(\phi _1\right) \cos \left(\phi _2\right)-J_3 \cos ^2\left(2 \theta _2\right) \sin \left(\phi _1\right) \sin \left(\phi _2\right), \\
	\dot \theta_2 &= J_e \cos \left(\phi _1\right) \cos \left(\phi _2\right)-J_3 \cos ^2\left(2 \theta _1\right) \sin \left(\phi _1\right) \sin \left(\phi _2\right),\\
    \dot \phi_1 &= J_o
	+\frac{\sin \left(2 \theta _1+2\theta _2\right) \left[J_e \sin \left(\phi _1\right) \cos \left(\phi _2\right)+J_3 \sin \left(\phi _2\right) \cos \left(\phi _1\right)\right]}{1+\cos \left(2 \theta _1\right) \cos \left(2 \theta _2\right)}, \\
	\dot \phi_2 &= J_o
	+\frac{\sin \left(2 \theta _1+2\theta _2\right) \left[J_e \sin \left(\phi _2\right) \cos \left(\phi _1\right)+J_3 \sin \left(\phi _1\right) \cos \left(\phi _2\right)\right]}{1+\cos \left(2 \theta _1\right) \cos \left(2 \theta _2\right)}.
  \end{aligned}   
 \end{equation}   
It is advantageous to reorganize the dynamics in terms of symmetric and antisymmetric combinations of the
variational parameters,
\begin{equation}
	\theta_{S/A} = \tfrac{1}{2}(\theta_1 \pm \theta_2), \qquad
	\phi_{S/A} = \tfrac{1}{2}(\phi_1 \pm \phi_2).
\end{equation}
In this basis, the equations of motion acquire a more compact and transparent form,
 \begin{equation}
 \begin{aligned}
    \dot{\theta}_S &= J_e\!\left[\cos^2(\phi_S)-\sin^2(\phi_A)\right] 
		-\tfrac{J_3}{2}\Big[\cos(4\theta_A)\cos(4\theta_S)+1\Big]\Big[\sin^2(\phi_S)-\sin^2(\phi_A)\Big], \\
	\dot{\theta}_A &= \tfrac{J_3}{4}\,\sin(4\theta_A)\sin(4\theta_S)\Big[\cos(2\phi_S)-\cos(2\phi_A)\Big], \\
    \dot{\phi}_S &= J_o + \frac{(J_e+J_3)\,\sin(4\theta_S)\,\sin(2\phi_S)}{2+\cos(4\theta_S)+\cos(4\theta_A)}, \\
	\dot{\phi}_A &= \frac{(J_e-J_3)\,\sin(4\theta_S)\,\sin(2\phi_A)}{2+\cos(4\theta_S)+\cos(4\theta_A)}.
     \end{aligned}
 \end{equation}

Several consistent reductions of this system are possible depending on the choice of initial conditions.  
First, if the system is initialized symmetrically with $\theta_1=\theta_2$, as we do in the main text, then $\theta_A \equiv 0$ is preserved 
by the dynamics, reducing the effective dimensionality to three variables $(\theta,\phi_S,\phi_A)$ with $\theta\equiv\theta_S$. 
The equations of motion then simplify to
\begin{equation}
\begin{aligned}
    \dot{\theta} &= J_e\!\left[\cos^2(\phi_S)-\sin^2(\phi_A)\right] 
		- J_3\cos^2(2\theta)\Big[\sin^2(\phi_S)-\sin^2(\phi_A)\Big], \\
    \dot{\phi}_S &= J_o + \frac{(J_e+J_3)\,\sin(4\theta)\,\sin(2\phi_S)}{3+\cos(4\theta)}, \\
	\dot{\phi}_A &= \frac{(J_e-J_3)\,\sin(4\theta)\,\sin(2\phi_A)}{3+\cos(4\theta)}.
\end{aligned}
\end{equation}

\begin{figure}[h]
    \centering   \includegraphics[width=0.3\linewidth]{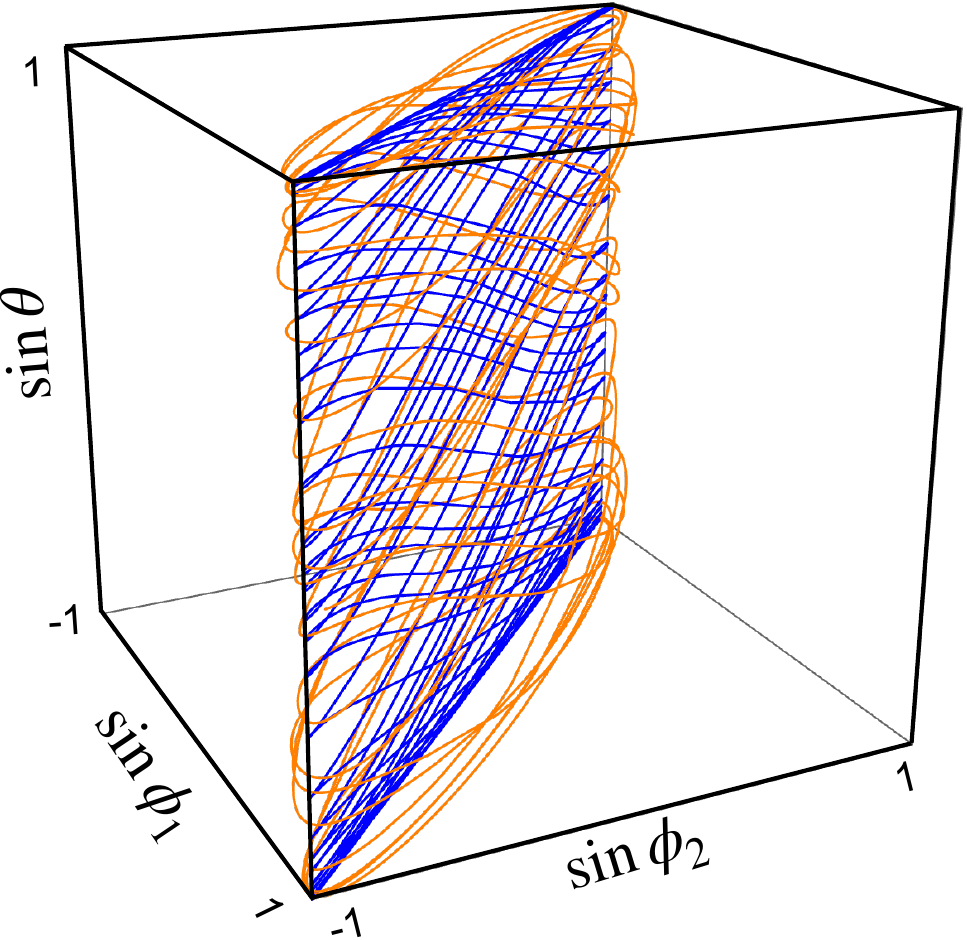}
    \caption{{\bf Impact of flow-invariant planes.}  
    Comparison between a TDVP trajectory constrained exactly to the invariant plane [initial condition $(\theta,\phi_1,\phi_2)=(0,0,0)$, blue] and one initialized slightly away from it [$(\theta,\phi_1,\phi_2)=(0,0,0.05\pi)$, orange]. On the invariant plane $\phi_1=\phi_2$, the TDVP equations reduce to two variables and produce strictly regular closed orbits, whereas trajectories launched near the plane initially follow the regular orbit but gradually drift into the surrounding phase space. 
}
    \label{fig:leakage}
\end{figure}

A further reduction occurs when the phases are initially aligned, $\phi_1=\phi_2\equiv\phi$, which enforces 
$\phi_A \equiv 0$. In this case the system closes on the two-dimensional manifold $(\theta,\phi)$,
\begin{equation}
\begin{aligned}
    \dot{\theta} &= J_e\cos^2\phi - J_3\cos^2(2\theta)\,\sin^2\phi, \\
    \dot{\phi}   &= J_o + \frac{(J_e+J_3)\,\sin(4\theta)\,\sin(2\phi)}{3+\cos(4\theta)}.
\end{aligned}
\end{equation}
This final reduction is particularly important: as a two-dimensional dynamical system it cannot exhibit chaos, 
which explains the absence of irregular trajectories along the corresponding boundary of phase space. Fig.~\ref{fig:leakage} illustrates this behavior: trajectories initialized exactly on the invariant plane remain confined to smooth, closed orbits, while those launched slightly away from it initially shadow the regular orbit before slowly drifting through the surrounding phase space, though they remain close to the invariant surface. 

\subsection{Exact rainbow scar point}

When $J_3=-J_e$, the model contains an exact solution associated with so-called rainbow scars~\cite{Langlett2021,Dong2023}. At this exact point, our ansatz faithfully captures the scar dynamics, where the TDVP equations of motion become:
\begin{equation}
\begin{aligned}
    \dot \theta_1 &= J_e \left[\cos \left(\phi _1\right) \cos \left(\phi _2\right) + \cos ^2\left(2 \theta _2\right) \sin \left(\phi _1\right) \sin \left(\phi _2\right)\right], \\
	\dot \theta_2 &= J_e \left[\cos \left(\phi _1\right) \cos \left(\phi _2\right) + \cos ^2\left(2 \theta _1\right) \sin \left(\phi _1\right) \sin \left(\phi _2\right) \right],\\
    \dot \phi_1 &= J_o
	+J_e\frac{\sin \left(2 \theta _1+2\theta _2\right)  \sin \left(\phi _1-\phi _2\right)}{1+\cos \left(2 \theta _1\right) \cos \left(2 \theta _2\right)}, \\
	\dot \phi_2 &= J_o
	-J_e\frac{\sin \left(2 \theta _1+2\theta _2\right) \sin \left(\phi _1-\phi _2\right)}{1+\cos \left(2 \theta _1\right) \cos \left(2 \theta _2\right)}.
\end{aligned}
\end{equation}
The scar initial state corresponds to the initialization $(\theta_1,\theta_2,\phi_1,\phi_2)=(0,0,0,\pi/2)$. In this case, the dynamics reduce further to
\begin{equation}
\begin{aligned}
    \dot \theta_1 &= \dot \theta_2 = 0,\\
    \dot \phi_1 &= \dot \phi_2 = J_o,
\end{aligned}
\end{equation}
which is exactly the rainbow scar dynamics~\cite{Dong2023}.

\begin{figure}
    \centering
    \includegraphics[width=\linewidth]{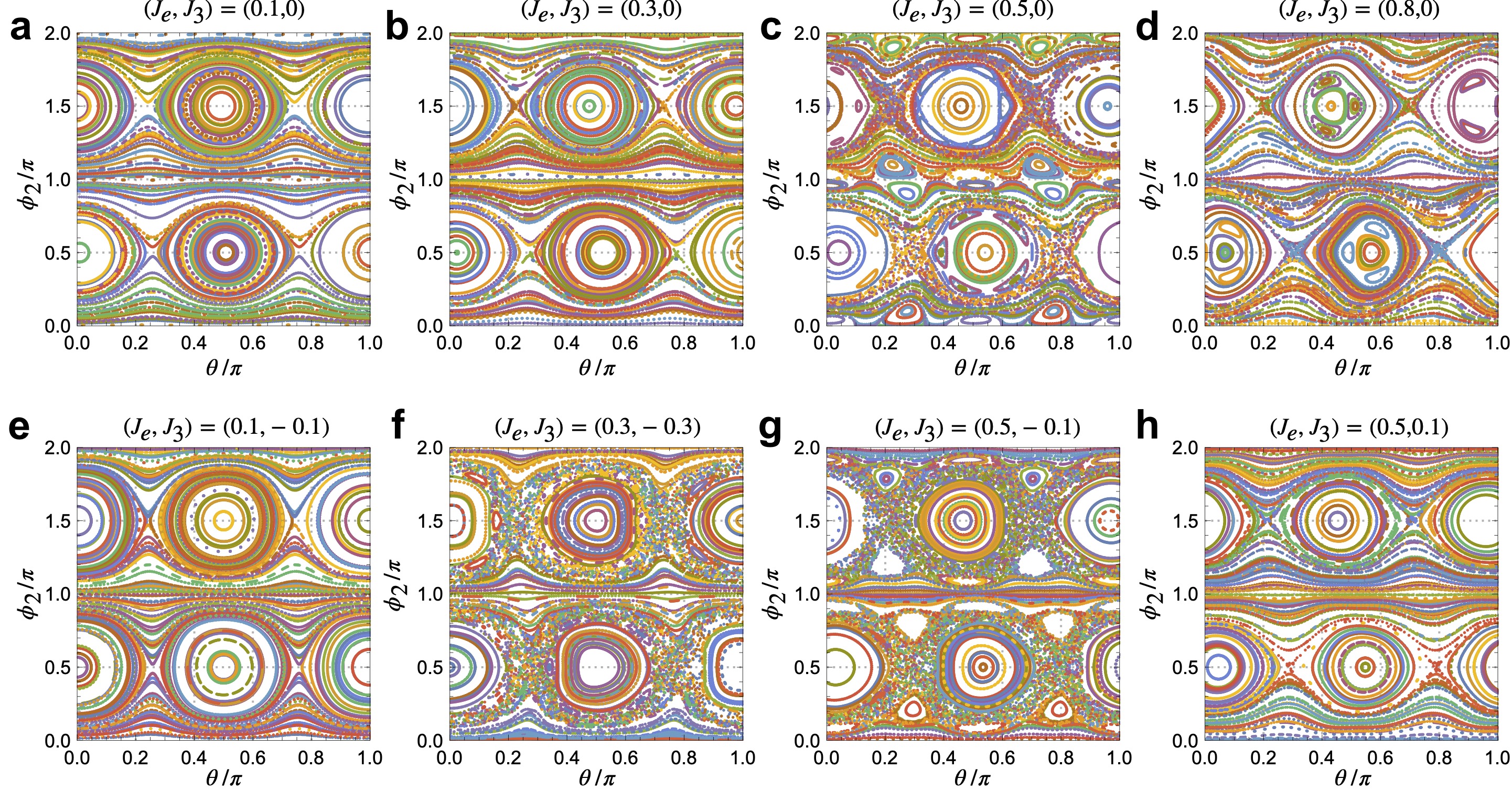}
    \caption{{\bf Evolution of the Poincar\'e section upon varying the circuit parameters.} Poincar\'e sections on the plane $\phi_1=0$, starting from the MPS ansatz with parameters $\theta_1=\theta_2=\theta,\phi_1=0$. We fix $J_o=1.0$, and the other two couplings are indicated in the title of each plot.}
    \label{fig:ps-1}
\end{figure}

\section{Poincar\'e sections for different couplings and under perturbations}

In this section, we study the evolution of the Poincar\'e sections for different choices of parameters and boundary conditions, as well as an experimentally-relevant perturbation.

First, we analyze the Poincar\'e section on the plane $\phi_1=0$, starting from initial conditions $\theta_1=\theta_2=\theta,\phi_1=0$. The resulting Poincar\'e sections are shown in Fig.~\ref{fig:ps-1}.
In Fig.~\ref{fig:ps-1}\textbf{a}-\textbf{d}, we show the noninteracting case where $J_3=0$; in Fig.~\ref{fig:ps-1}\textbf{e}-\textbf{f}, we fix $J_e=-J_3$; in Fig.~\ref{fig:ps-1}\textbf{g}-\textbf{h}, we choose generic couplings. In all three cases, we observe both mixed-phase and regular patterns in the phase space. It is important to note that the details of the phase space profile are sensitive to the difference in the strength of $J_e$ and $J_3$. When $|J_e-J_3|<0.5$, the profile is simpler and consists of a couple of large islands, while otherwise these islands develop substructure and smaller tori also appear outside of these islands.

\begin{figure}
    \centering
    \includegraphics[width=\linewidth]{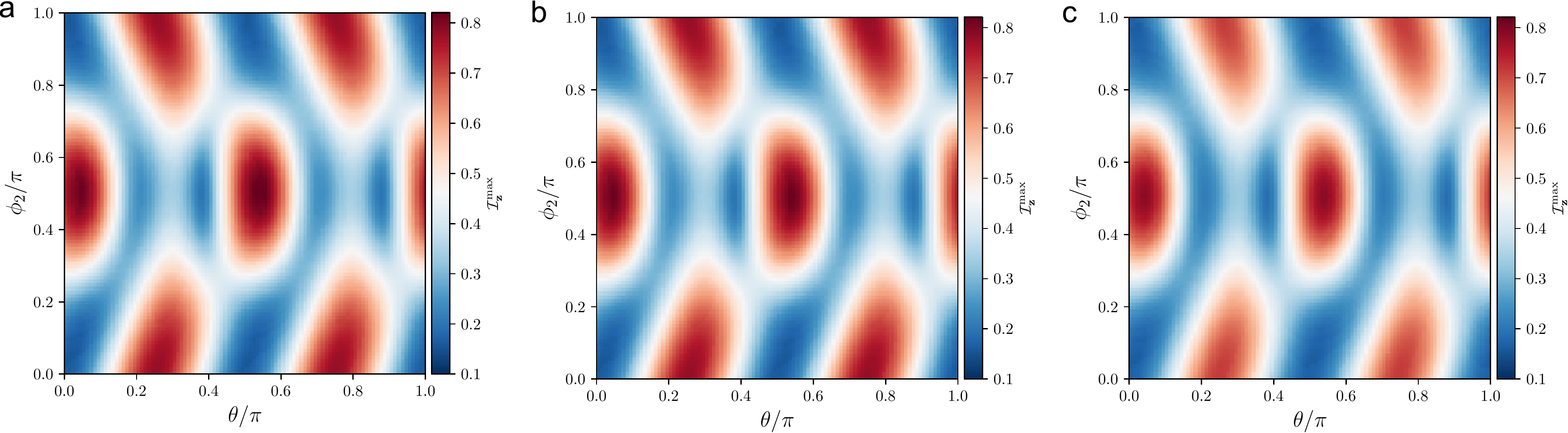}
    \caption{{\bf Insensitivity of the Poincar\'e section to perturbations and type of boundary conditions.} \textbf{a,b} First revival peak in local observables from exact diagonalization, for the Hamiltonian under \textbf{a}  open boundary conditions and \textbf{b} periodic boundary conditions. \textbf{c} First revival peak for the Hamiltonian with diagonal perturbation in Eq.~\eqref{suppeq:dperturbation} with $J_d=0.1 J_o$. All simulations are performed using exact diagonalization of a 16-qubit system.}
    \label{fig:ED}
\end{figure}

Next, we analyze the effect of boundary conditions. We perform exact diagonalization of a $16$-qubit chain under both open and periodic boundary conditions. As shown in Fig.~\ref{fig:ED}\textbf{a},\textbf{b}, the qualitative structure of the revival map is nearly identical in the two cases, indicating that the coexistence of regular and chaotic regions is not an artifact of the boundary condition.

Finally, we explore the effect of an experimentally-relevant diagonal perturbation:
\begin{equation}\label{suppeq:dperturbation}
    \hat V_d = -J_d \sum_j \left(\hat\sigma^+_j \hat\sigma^-_{j+2} + \hat\sigma^-_j \hat\sigma^+_{j+2}\right),
\end{equation}
which introduces weak next-nearest-neighbor exchange along the ladder. In superconducting-qubit implementations, such diagonal couplings naturally arise from residual capacitive or inductive crosstalk between nonadjacent qubits, or from higher-order virtual processes mediated by the tunable couplers. These mechanisms effectively generate weak flip-flop interactions between qubits on opposite legs of the ladder, corresponding to the diagonal links in the model. We fix $J_d = 0.1J_o$---the scale of perturbation typically observed experimentally---and compute the first revival peak of the local observable. As shown in Fig.~\ref{fig:ED}\textbf{c}, the resulting phase-space pattern remains qualitatively unchanged, confirming that the mixed-phase structure is robust against these types of hardware imperfections. This resilience highlights that the observed coexistence of regular and chaotic trajectories is an intrinsic feature of the variational dynamics, rather than a fine-tuned property of the idealized Hamiltonian.

\begin{figure}
    \centering
    \includegraphics[width=0.8\linewidth]{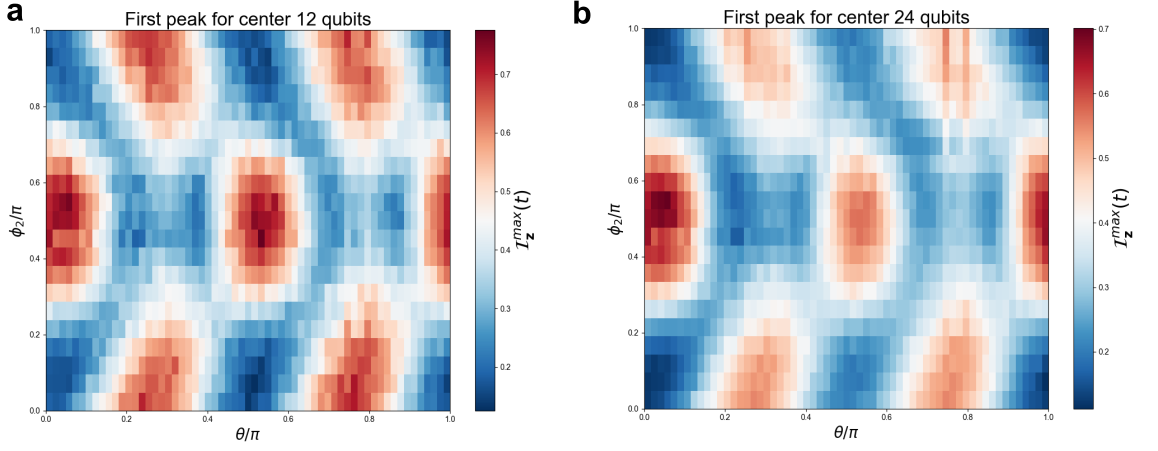}
    \caption{\textbf{Experimental observation of the first revival peak via imbalance in the 24-qubit system}. To demonstrate the robustness of mixed phase space, we contrast the imbalance measured on (a) the central 12 qubits, and (b) all 24 qubits of the ladder. The larger fluctuations in panel (b) are attributed to the boundary effects. The coupling parameters are $J_o/2\pi=5$ MHz, $J_e/2\pi=3$ MHz, and $J_3/2\pi=0.25$ MHz.
    }
    \label{fig:exp_ED}
\end{figure}

The robustness of the mixed phase-space structure is further illustrated in Fig.~\ref{fig:exp_ED}, which compares the imbalance-based Poincar\'e sections obtained from the central 12-qubit block with those extracted from all 24 qubits. While the full-system measurement reproduces the same qualitative features, including the location of the regular islands and the surrounding chaotic regions, the boundary spins introduce additional fluctuations. For this reason, the central subsystem provides a cleaner and more reliable experimental probe of the underlying phase-space structure, and is used throughout the main text.

\begin{figure}[tb]
    \centering
    \includegraphics[width=0.7\linewidth]{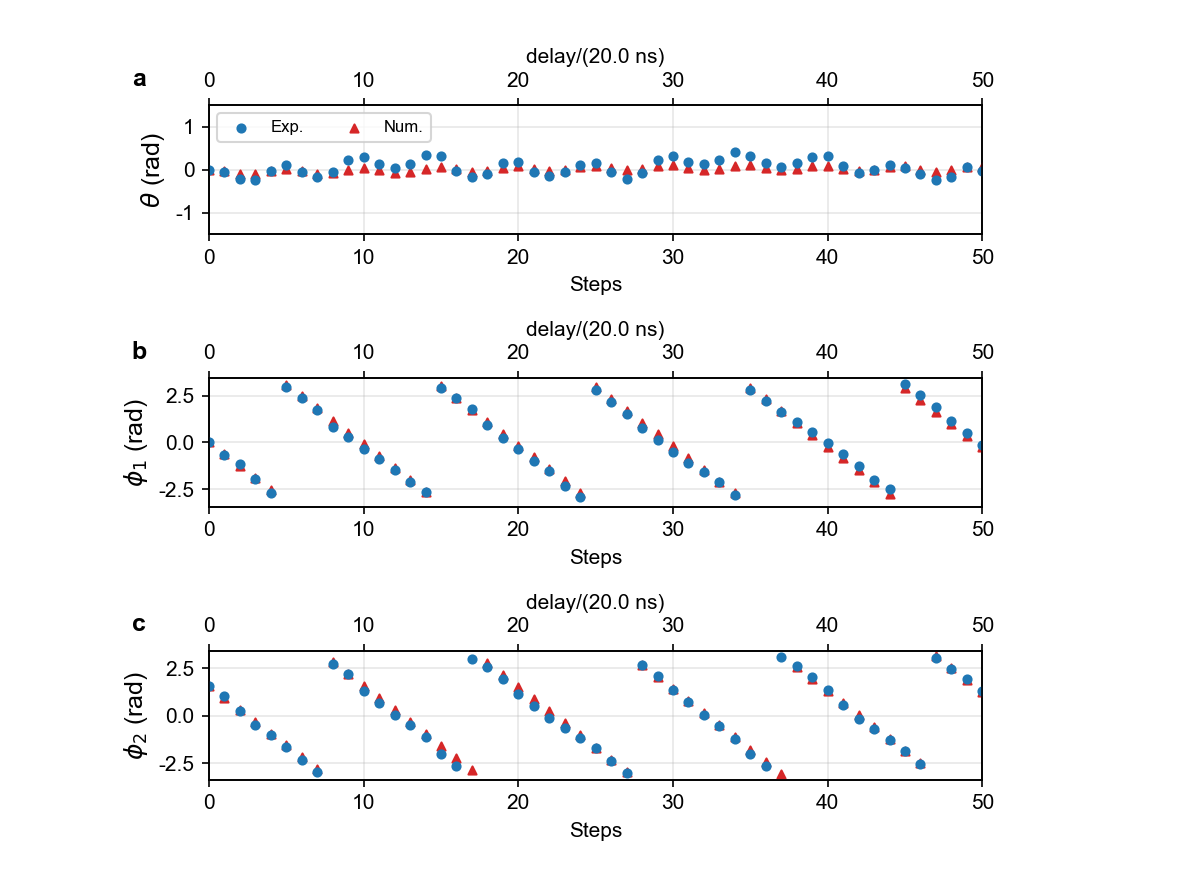}
    \caption{\textbf{The dynamics of the hybrid feedback dynamics compared to the numerical results} 
    (a–c) Evolution of the parameters ($\theta$, $\phi_1$, $\phi_2$) under the hybrid feedback protocol, starting from the matrix product state $(0,0,0.5\pi)$ with a fixed delay of 20 ns. The numerical results are shown for comparison. The remarkable agreement between both methods---even after a large number of steps---demonstrates the effectiveness of the hybrid feedback approach in capturing the essential dynamics of the system.}
    \label{fig:dynamics_scarfinder}
\end{figure}

\section{Convergence of the hybrid protocol}

Step 4 of the algorithm in the main text can be viewed as maximizing the fidelity, $\max_{\bm z}\, |\langle\psi(\bm z)|\psi(t)\rangle|^2$, i.e., the overlap of the true time-evolved state with the closest variational state on $\mathcal{M}$. This step therefore realizes a discrete projection of the quantum trajectory back onto $\mathcal{M}$, with a clear physical meaning. A generic many-body state contains both a low-entanglement component $|\psi_\mathcal{M}\rangle$ and a highly-entangled residual $|\psi_\perp\rangle$, such that $|\psi(t)\rangle = |\psi_\mathcal{M}(t)\rangle + |\psi_\perp(t)\rangle$. 
Under chaotic evolution, $|\psi_\perp\rangle$ rapidly accumulates entanglement, causing destructive interference and dephasing in observable quantities. In contrast, $|\psi_\mathcal{M}\rangle$ evolves quasi-classically, following the TDVP flow and therefore preserves coherent oscillations over long times. The hybrid feedback effectively enforces the iterative map $ |\psi^{(n+1)}\rangle = 
    \mathcal{P}_{\mathcal{M}}\, \hat U(\Delta t)\,|\psi^{(n)}\rangle$. Repeated application of this map suppresses the entropy-generating component $|\psi_\perp\rangle$ while amplifying the smooth, quasi-regular motion within $\mathcal{M}$. When the feedback loop converges, it identifies a self-consistent fixed point
\begin{equation}
    e^{-i\hat{H}T}\,|\psi^*\rangle \simeq e^{i\varphi}\,|\psi^*\rangle,
\end{equation}
corresponding to a (quasi-)periodic trajectory on the variational manifold. This orbit represents the most stable initial state accessible within the ansatz -- a trajectory that minimizes variational leakage and exhibits maximally sustained coherence under repeated evolution and projection.
In this sense, the hybrid feedback loop acts as a variational distillation process: each iteration filters out unstable, rapidly thermalizing motion, leaving behind the coherent low-dimensional dynamics.  The corresponding experimental \textsc{ScarFinder} trajectory, obtained from the same update rule, is shown in Fig.~\ref{fig:dynamics_scarfinder} for comparison.

\begin{figure}[tb]
    \centering
    \includegraphics[width=0.45\linewidth]{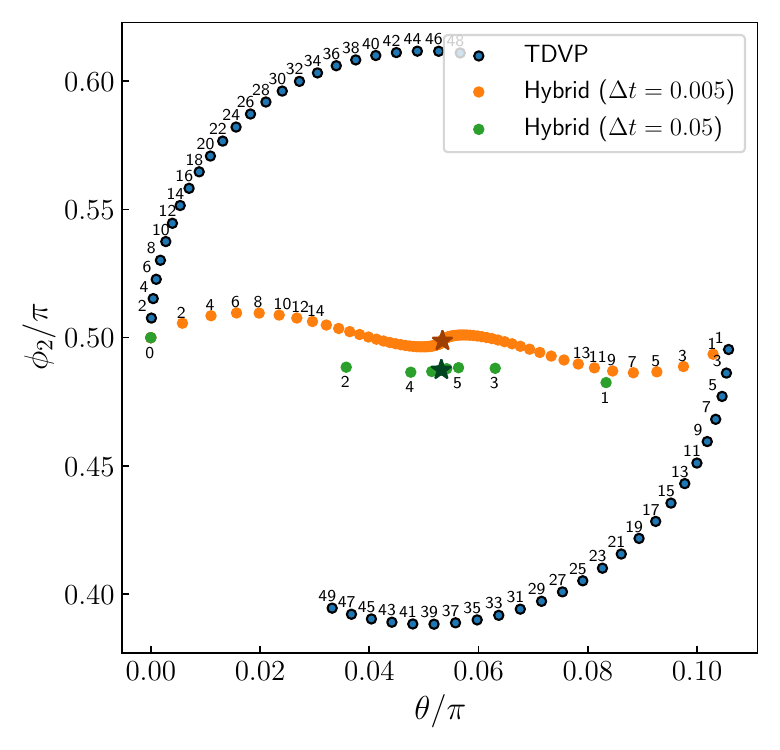}
    \caption{\textbf{Comparison between TDVP and hybrid feedback trajectories.}
    Poincar\'e section of the variational dynamics obtained from TDVP (blue points) and from the hybrid feedback algorithm (orange and green points). For small feedback step size $\Delta t$, each iteration of the hybrid protocol closely approximates continuous TDVP evolution, producing a discrete trajectory in the variational parameter space. Whenever the trajectory crosses the Poincaré plane, a point is recorded, forming the spiral patterns shown. The spiral illustrates how the iterative feedback dynamics gradually converge toward the center of the regular island, corresponding to a more stable periodic orbit. Different values of $\Delta t$ lead to distinct convergence rates, and the final converged orbit is slightly deformed depending on $\Delta t$, reflecting the finite-step discretization of the continuous TDVP flow.}
    \label{fig:converge}
\end{figure}

The hybrid feedback protocol can be viewed as a discrete approximation of the continuous TDVP flow, where each feedback cycle represents a finite step of duration $\Delta t$. In the limit $\Delta t \to 0$, the projection–evolution sequence becomes equivalent to the infinitesimal TDVP evolution, and the hybrid trajectory coincides with the continuous variational dynamics. For finite $\Delta t$, however, the discrete update introduces a small systematic deviation from the exact TDVP path, which acts as an effective dissipative correction. This correction gradually steers the system toward the nearest stable periodic orbit of the TDVP manifold.
Fig.~\ref{fig:converge} visualizes this process on the Poincar\'e section of the variational manifold. The blue points represent intersections of TDVP trajectories with the chosen plane, forming closed regular orbits. The colored spirals correspond to successive iterations of the hybrid protocol for different feedback step sizes $\Delta t$. Each iteration approximates the TDVP flow but, due to the finite-step discretization, the trajectory slowly spirals inward toward the center of the regular island. The center corresponds to a locally most stable orbit, characterized by minimal leakage from the variational manifold and maximal revival fidelity.

The rate of convergence depends sensitively on the choice of $\Delta t$. Smaller $\Delta t$ provides a more accurate approximation of the continuous TDVP evolution, leading to slower but smoother convergence toward the fixed point. Larger $\Delta t$ accelerates the approach but also amplifies discretization errors, causing the final orbit to deviate slightly from the ideal TDVP trajectory. Consequently, the steady-state orbit obtained after many iterations is not identical for different $\Delta t$, but rather represents a deformed version of the underlying continuous orbit. This deformation vanishes smoothly in the limit $\Delta t \to 0$, confirming that the hybrid feedback protocol continuously interpolates between a dissipative discrete map and the conservative TDVP flow.

\begin{figure}[tb]
    \centering
    \includegraphics[width=0.6\linewidth]{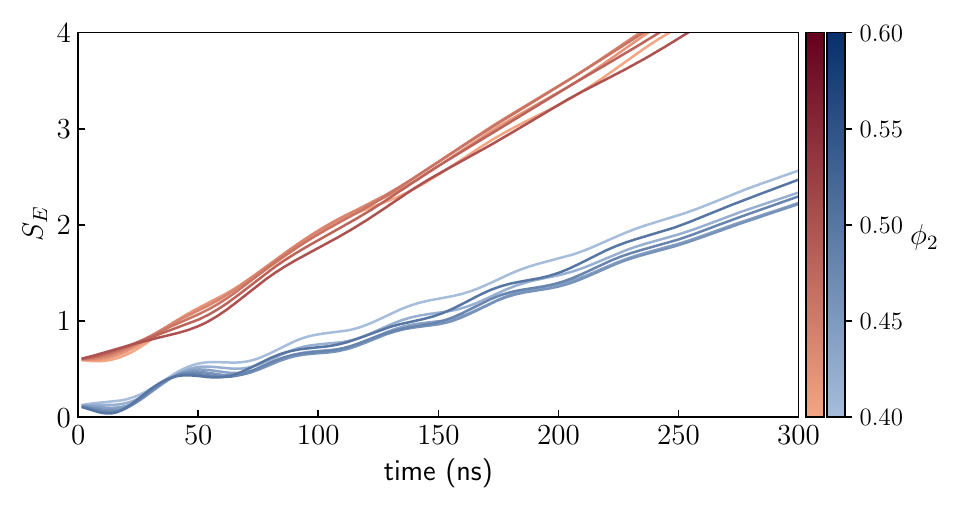}
    \caption{
    \textbf{Signatures of quantum many-body mixed phase space in the thermodynamic limit.} Time evolution of the bipartite entanglement entropy $S_E$ for two families of initial states within the shallow-circuit manifold. The blue curves correspond to states $|\xi(\phi_2)\rangle \equiv |\psi(0.05\pi, 0, \phi_2)\rangle$  located in the regular island of phase space in Fig.~1{\bf d} of the main text, while the orange curves correspond to chaotic states $|\eta(\phi_2)\rangle \equiv |\psi(0.32\pi, 0, \phi_2)\rangle$. The angle $\phi_2$ is varied in the interval  $[0.4\pi,0.6\pi]$, as indicated by the color gradient. The distinct slopes of entropy growth reflect markedly different dynamical regimes (slow, oscillatory evolution versus rapid entanglement spreading), consistent with our Poincar\'e section for finite $N$ in Fig.~2 of the main text. The simulations were performed using iTEBD on an iMPS representation with maximum bond dimension $\chi = 256$. 
    }
    \label{fig:mixed_phase_ee}
\end{figure}

\section{Quantum many-body mixed phase space in the thermodynamic limit}

While our study of mixed phase space in the main text focused on finite systems, here we demonstrate that the interacting SSH ladder exhibits similar ergodicity-breaking features even in the thermodynamic limit, $N{\to}\infty$. As explained in Sec.~\ref{apx:mps_ansatz}, the shallow circuit ansatz states from the main text can be conveniently expressed as infinite matrix-product states (iMPS), which we denote by $|\psi(\theta, \phi_1, \phi_2)\rangle$. Using time-evolving block decimation (TEBD)~\cite{Vidal07}, we analyze the evolution of bipartite entanglement entropy, $S_E=-\mathrm{tr}\rho_A\ln \rho_A$, where $\rho_A$ is the reduced density matrix of one half of an infinite chain. We fix $\phi_1=0$ and consider two representative families of initial conditions,
$|\xi(\phi_2)\rangle$ belonging to the regular island and 
$|\eta(\phi_2)\rangle$ in the chaotic region of Fig.~1{\bf d} of the main text, as we vary $\phi_2 \in [0.4\pi, 0.6\pi]$.  
Although these states share the same energy expectation, $\langle \hat{H} \rangle = 0$, which places them at the center of the many-body spectrum, their entanglement dynamics differ qualitatively, as shown in Fig.~\ref{fig:mixed_phase_ee}.  

According to the ETH, states at similar energies should exhibit similar thermalization dynamics and entropy growth rates. By contrast,  Fig.~\ref{fig:mixed_phase_ee} reveals two clearly distinct dynamical responses: states belonging to the $\xi$ family show slow, oscillatory growth characteristic of regular trajectories, while the $\eta$ family displays rapid, almost linear entanglement growth indicative of chaotic dynamics. Similar behavior is seen in the relaxation behavior of local observables. This coexistence of distinct dynamical regimes at the same energy density is a weak violation of ETH and points to the emergence of mixed phase space structure within the variational manifold, consistent with our finite-$N$ results in Fig.~2 of the main text.

\section{Refined Poincar\'e section analysis}

\begin{figure}[htb]
    \centering   \includegraphics[width=0.8\linewidth]{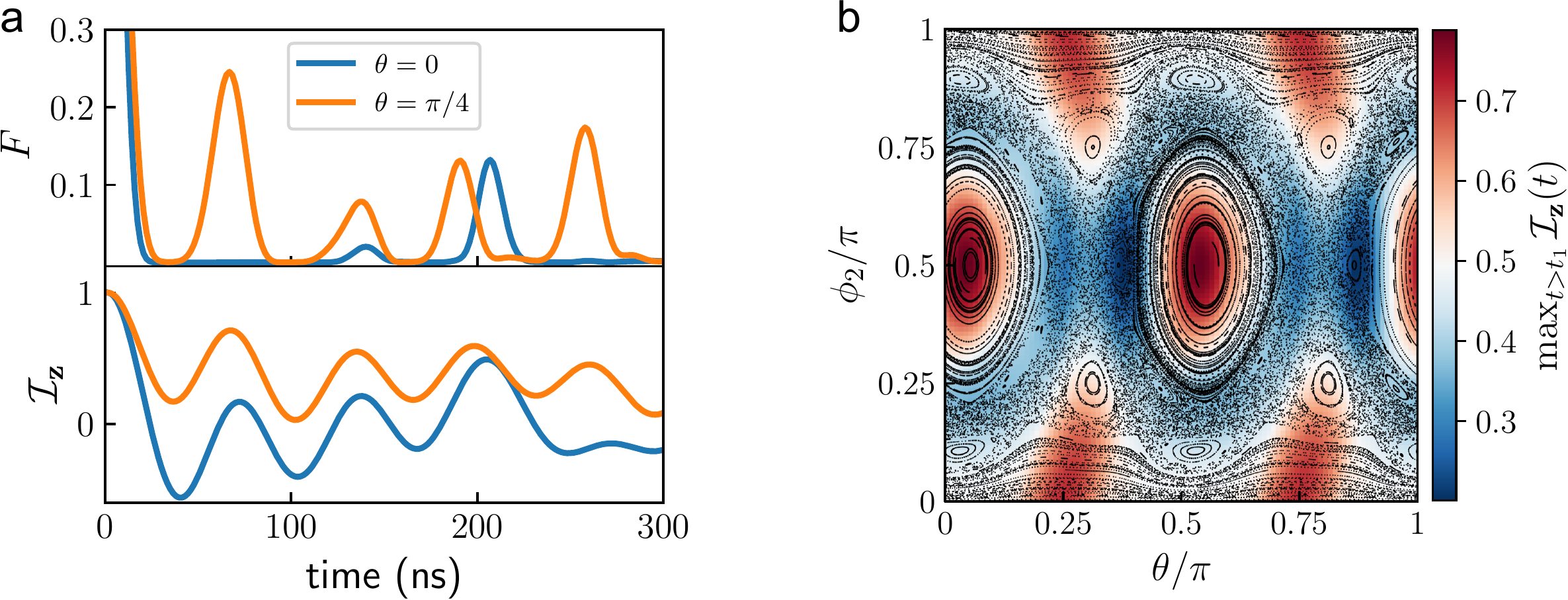}
    \caption{{\bf Refined Poincar\'e section mapping.}
    \textbf{a} Time evolution of fidelity (top) and imbalance (bottom) for two representative initial states with $\phi_1=\phi_2=0$, $\theta=0$ (blue) and $\theta=\pi/4$ (orange). The two observables display qualitatively similar revival patterns. For $\theta=\pi/4$, the strongest revival occurs at the first peak, whereas for $\theta=0$, the maximal signal appears only at the third revival.  
    \textbf{b} Maximal revival amplitude of the imbalance, $\max_{t>t_1}\mathcal{I}_{\bm z}(t)$, mapped as a function of $(\theta,\phi_2)$ and overlaid with the TDVP Poincar\'e section (black dots). 
    The regions of high revival amplitude align closely with the TDVP regular islands, while the low-revival areas correspond to the chaotic sea. Compared to $\mathcal{I}_{\bm z}^\mathrm{max}$, this refined measurement achieves an improved agreement with the TDVP Poincar\'e section. Numerical results are obtained by exact diagonalization of an $L=16$ spin chain with periodic boundary conditions.
    }
    \label{fig:refined-poincare}
\end{figure}

As mentioned in the main text, the Poincar\'e sections inferred from imbalance measurements can show small discrepancies from the TDVP predictions, e.g., near the upper and lower boundaries of the phase space in Fig.~2 of the main text. We can gain intuition about these discrepancies if we restrict to symmetric initial conditions, $\phi_1=\phi_2=0$, for which the dynamics reduce to a two-dimensional submanifold. Since two-dimensional continuous dynamical systems cannot display deterministic chaos due to the Poincar\'e–Bendixson theorem~\cite{StrogatzBook}, this reduction implies strictly regular, although not necessarily periodic, trajectories. Figure~\ref{fig:refined-poincare}{\bf a} illustrates this point more explicitly for fixed $\phi_1=\phi_2=0$ and two states with $\theta=0$ and $\theta=\pi/4$, which exhibit qualitatively different dynamical behavior, even though both lie within the regular region of the TDVP Poincar\'e section. In terms of fidelity $F(t)\equiv|\langle\psi(0)|\psi(t)\rangle|^2$ (top panel of Fig.~\ref{fig:refined-poincare}{\bf a}) the $\theta=\pi/4$ state reaches its maximal value at the first revival, whereas for $\theta=0$ the fidelity continues to increase over several revivals, with the largest peak appearing at the third revival. Hence, a diagnostic based solely on the first peak would  incorrectly classify the $\theta=0$ dynamics as irregular, despite the clear presence of oscillations. The same trend is evident in the imbalance dynamics (bottom panel of Fig.~\ref{fig:refined-poincare}{\bf a}): for $\theta=\pi/4$, the peak amplitude of $\mathcal{I}_{\bm z}(t)$ decays gradually, while for $\theta=0$, successive revivals grow in strength, mirroring the behavior observed in fidelity. Consequently, restricting the analysis to the first revival underestimates the stability of these trajectories and misrepresents the corresponding regions in the imbalance-based Poincar\'e map.

To overcome this limitation, we refine our diagnostic by extending the observation window in time and taking the \emph{maximal} revival amplitude, regardless of when it occurs, $\max_{t>t_1}\mathcal{I}_{\bm z}(t)$,  where $t_1$ denotes the time of the first minimum following initialization. This refined definition captures trajectories whose strongest coherence reemerges at later times, providing a more faithful correspondence to the underlying TDVP dynamics. The resulting refined Poincar\'e section, shown in Fig.~\ref{fig:refined-poincare}{\bf b}, exhibits markedly improved agreement with the TDVP prediction, correctly recovering the high-stability regions near the top and bottom of the parameter space that were previously underestimated. This demonstrates that the discrepancies stem mainly from focusing on the first revival and ignoring the full periodic structure of the imbalance dynamics, rather than from fundamental deficiencies of the imbalance-based measurement protocol.

\end{document}